\documentclass{aa}  
\usepackage{graphics,graphicx}
\usepackage{xcolor}
\usepackage{enumitem}
\usepackage{float}
\usepackage{caption}
\usepackage{subcaption}

\usepackage{txfonts}
\usepackage{longtable}
\usepackage[unicode=true,pdfusetitle,
 bookmarks=true,bookmarksnumbered=false,bookmarksopen=false,
 breaklinks=false,pdfborder={0 0 0},pdfborderstyle={},backref=false,colorlinks=true]
 {hyperref}
 
\hypersetup{
 urlcolor=blue,citecolor=blue}

\begin{document}  
   \title{Measuring the ICM velocity structure within the A3266 galaxy cluster} 

   \author{E. Gatuzz\inst{1},   
           J. Sanders\inst{1}, 
           A. Liu\inst{1}, 
           A. Fabian\inst{2}, 
           C. Pinto\inst{3}, 
           H. Russell\inst{4},   
           D. Eckert\inst{5}, \\ 
           S. Walker\inst{6}
           J. ZuHone\inst{7}   \and
           R. Mohapatra\inst{8}
          } 
 
   \institute{Max-Planck-Institut f\"ur extraterrestrische Physik, Gie{\ss}enbachstra{\ss}e 1, 85748 Garching, Germany\\
              \email{egatuzz@mpe.mpg.de}
         \and
             Institute of Astronomy, Madingley Road, Cambridge CB3 0HA, UK
         \and
              INAF - IASF Palermo, Via U. La Malfa 153, I-90146 Palermo, Italy 
          \and
             School of Physics \& Astronomy, University of Nottingham, University Park, Nottingham NG7 2RD, UK
          \and
             Department of Astronomy, University of Geneva, Ch. d\rq Ecogia 16, CH-1290 Versoix, Switzerland
          \and
             Department of Physics and Astronomy, University of Alabama in Huntsville, Huntsville, AL 35899, USA
          \and
             Harvard-Smithsonian Center for Astrophysics, 60 Garden Street, Cambridge, MA, 02138, USA         
          \and
             Department of Astrophysical Sciences, Princeton University, NJ 08544, USA                       
             }
 
   \date{Received XXX; accepted YYY} 
 
  \abstract  
{We present a detailed analysis of the velocity structure of the hot intracluster medium (ICM) within the A3266 galaxy cluster, including new observations taken between June and November 2023. 
Firstly, morphological structures within the galaxy cluster were examined using a Gaussian Gradient Magnitude (GGM) and adaptively smoothed GGM filter applied to the EPIC-pn X-ray image. 
Then, we applied a novel {\it XMM-Newton} EPIC-pn energy scale calibration, which uses instrumental Cu K$\alpha$ as reference for the line emission, to measure line-of-sight velocities of the hot gas within the system.
This approach enabled us to create two-dimensional projected maps for velocity, temperature, and metallicity, showing that the hot gas displays a redshifted systemic velocity relative to the cluster redshift across all fields of view. 
Further analysis of the velocity distribution through non-overlapping circular regions demonstrated consistent redshifted velocities extending up to 1125 kpc from the cluster core. 
Additionally, the velocity distribution was assessed along regions following surface brightness discontinuities, where we observed redshifted velocities in all regions, with the largest velocities reaching $768 \pm 284$ km/s.
Moreover, we computed the velocity Probability Density Function (PDF) from the velocity map. 
We applied a normality test, finding that the PDF adheres to an unimodal normal distribution consistent with theoretical predictions.  
Lastly, we computed a velocity structure function (VSF) for this system using the measured line-of-sight velocities.
These insights advance our understanding of the dynamic processes within the A3266 galaxy cluster and contribute to our broader knowledge of ICM behavior in merging galaxy clusters. 
}

\titlerunning{ICM velocity structure within A3266}
\authorrunning{Gatuzz et al.}
\maketitle

\section{Introduction}\label{sec_int} 
The intracluster medium (ICM) is a hot, ionized gas that fills the space between galaxies in a galaxy cluster, emitting X-rays due to its high temperature of several million kelvins.
Measuring the velocities of the intracluster medium (ICM) is crucial for several reasons. 
Turbulent motions provide additional pressure support, particularly at large radii, affecting cluster mass estimates and calculations of hydrostatic equilibrium \citep{lau09}. 
Simulations predict a close connection between entropy, temperature, density, and velocity power spectra, which should be tested \citep{gas14}. 
Velocity measurements can help constrain AGN feedback models, as the distribution of energy from AGN feedback within the cluster depends on the balance between turbulence and sound waves or shocks. 
Additionally, these motions contribute to the transport of metals within the ICM through processes such as the uplift and sloshing caused by AGN \citep[e.g.,][]{sim08}. 
The microphysics of the ICM, including viscosity, can also be probed by measuring velocities \citep{zuh16,zuh18}. 
Furthermore, velocity measurements can directly detect the sloshing of gas in cold fronts, which can persist for several Gyr \citep{asc06}.

The {\it Hitomi} observatory directly measured random and bulk motions in the ICM using the Fe-K emission lines, thanks to its high spectral resolution microcalorimeter SXS X-ray detector. 
In the core of the Perseus cluster, it measured a bulk flow gradient of $150$ km/s across $60$ kpc of the cluster core and a line-of-sight velocity dispersion of $164 \pm 10$ km/s between radii of $30-60$ kpc \citep{hit16}. 
These results, obtained for a very limited spatial region, showed that the Perseus core is not strongly turbulent despite the apparent impact of the AGN and its jets on the surrounding ICM and the merger-driven sloshing motions. 
Such a low level of turbulence argues against it providing distributed heating, although sound waves or shocks could plausibly do this as well as heating from merging activity \citep{fab17}. 
Unfortunately, no further measurements were made in other clusters or different regions of Perseus due to the loss of {\it Hitomi}, although the new {\it XRISM} mission is expected to extend the Fe K velocity measurements in more clusters.

{\it Suzaku} also obtained velocity measurements in several systems using the Fe-K line, thanks to its accurate calibration. 
\citet{tam14} placed upper limits on relative velocities of 300 km/s over scales of 400 kpc in Perseus. 
\citet{ota16} examined several clusters with {\it Suzaku}, though systematic errors from the calibration were likely around 300 km/s, and its point spread function (PSF) was large. 
Other methods of measuring velocities include using {\it XMM-Newton} RGS spectra to measure line widths \citep[e.g., ][]{pin15} and examining resonant scattering \citep[e.g.,][]{aho16}, both limited to the cluster core. 
Indirect measurements of velocity structure in clusters include analyzing the power spectrum of density fluctuations and linking this via simulations to the velocity spectrum \citep[e.g.,][]{zhu14}, or examining the magnitude of thermodynamic perturbations \citep{hof16}. 
However, these methods are model-dependent.

 \begin{figure}
        \centering 
            \includegraphics[width=0.48\textwidth]{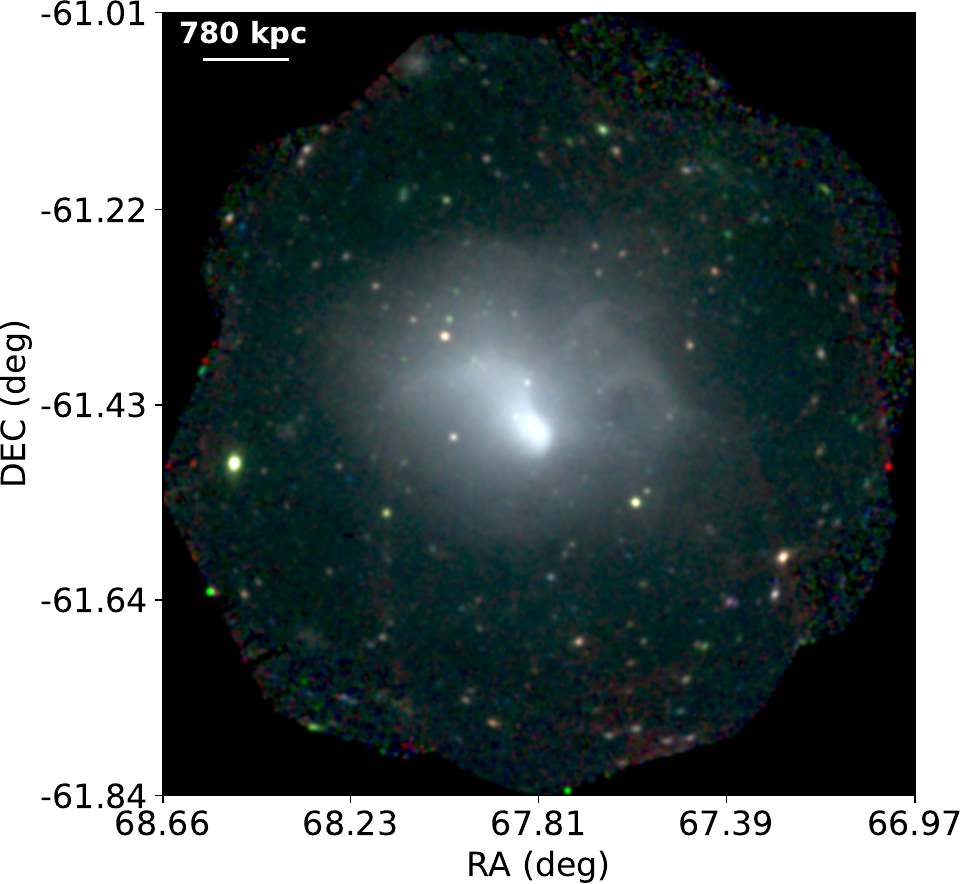}\\         
            \includegraphics[width=0.48\textwidth]{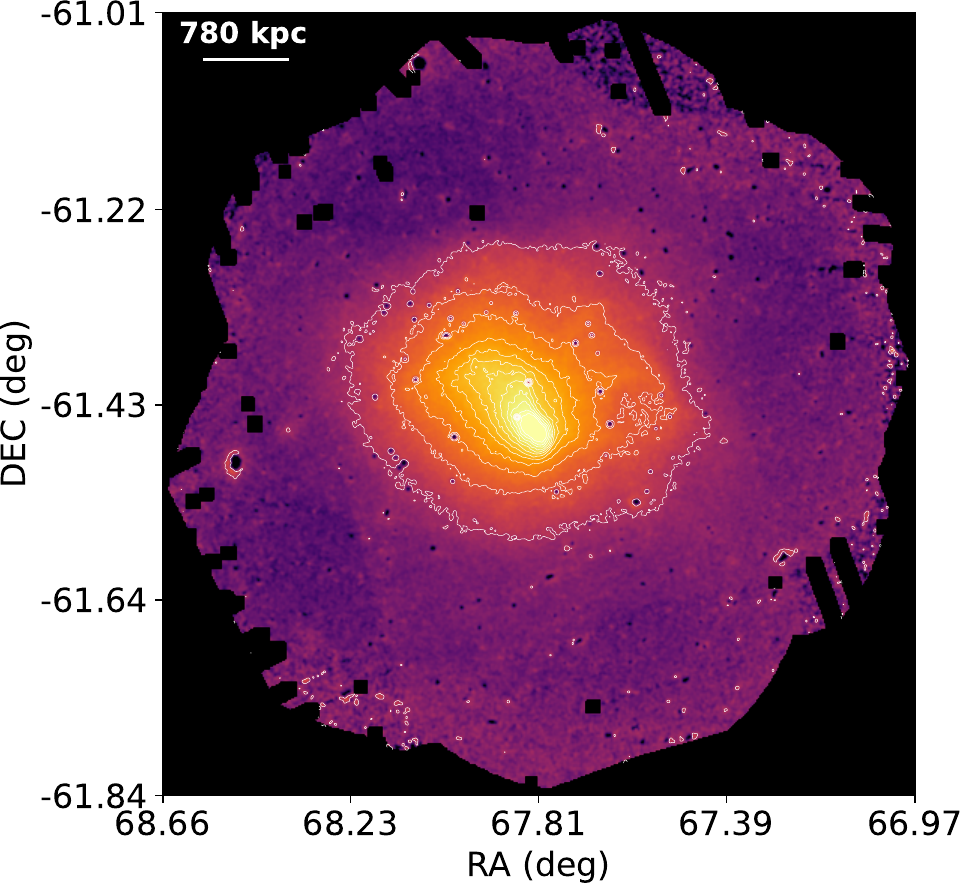}  
        \caption{\emph{Top panel:} RGB image of A3266. The bands used are 0.2-0.5~keV (R), 0.5-1.0~keV (G) and 1.0-2.0~keV (B). We applied a $\sigma=1$ Gaussian smoothing to the exposure corrected and background subtracted image. \emph{Bottom panel:} X-ray surface brightness exposure corrected and background subtracted of the A3266 cluster in the 0.5 to 9.25 keV energy range. The point-like sources were excluded in the analysis. The 10-level contours of the X-ray image are shown in white.  } \label{fig_xray_maps1} 
\end{figure}

\citet[][hereafter SAN+20]{san20} present a novel technique using background X-ray lines observed in the spectra of the {\it XMM-Newton} EPIC-pn detector to calibrate the absolute energy scale of the detector to better than $150$ km/s at Fe-K. 
Using this technique, SAN+20 mapped the bulk velocity distribution of the ICM over a large fraction of the Perseus and Coma clusters. 
For the Perseus cluster, they detected evidence of sloshing associated with a cold front. 
In the Coma cluster, SAN+20 found that the velocity of the gas closely matches the optical velocities of the two central galaxies, NGC 4874 and NGC 4889. 
However, the velocity structure near the cluster center was not studied for either source due to the lack of offset observations covering the central region.

In recent years, four large {\it XMM-Newton} observation programs were approved to observe the Virgo, Centaurus, and Ophiuchus clusters and apply this method. 
Using this technique, we have obtained accurate velocity measurements with uncertainties down to $\Delta v \sim 100$ km/s. 
The analysis of the Virgo cluster shows signatures of both AGN outflows and gas sloshing in the velocity field of the cluster \citep{gat22a}. 
For the Centaurus cluster, we have found (a) an overall gradient in velocities at large radii, (b) a complex velocity profile near the cluster center, and (c) changes in velocities along the cold fronts \citep{gat22b}. 
In the Ophiuchus cluster, we have identified a large redshifted-blueshifted interface located approximately 150 kpc east of the cluster core \citep{gat23a}. 
This interface, which shows signs of a lack of metal mixing, is traced by discontinuities in the X-ray surface brightness. 
These large observational campaigns also allow us to measure the velocity structure functions (VSF) within the systems \citep{gat23c}. 
VSFs constitute a useful diagnostic tool to study the turbulence in a medium, since they represent the variation of velocity with scale \citep{fed10,fed21,set23}.
Furthermore, a detailed analysis of the metallicity distribution in the context of the velocity structure has been carried out for these sources \citep{gat23b,gat23d,gat23e}.

\begin{table} 
\scriptsize
\caption{\label{tab_obsids}{\it XMM-Newton} observations of the A3266 cluster.}
\centering 
\begin{tabular}{cccccccc}   
\hline
ObsID & RA & DEC & Date & Exposure \\
 &&&Start-time& (ks)\\
\hline
0105262001&04:31:37.00&-61:34:02.0&2000-09-23&9.3\\
0105261101&04:32:35.49&-61:29:01.0&2000-09-25&15.0\\
0105262101&04:32:35.49&-61:29:01.0&2000-09-25&9.7\\
0105261001&04:30:13.30&-61:21:02.0&2000-09-27&15.4\\
0105262201&04:30:13.30&-61:21:02.0&2000-09-27&9.2\\
0105260701&04:32:26.70&-61:18:46.0&2000-10-01&22.1\\
0105262301&04:31:15.80&-61:15:33.0&2000-10-03&34.1\\
0105260901&04:30:23.60&-61:30:32.0&2000-10-09&26.0\\
0105260801&04:31:15.80&-61:15:33.0&2000-10-11&21.8\\
0105262501&04:31:37.00&-61:34:02.0&2003-03-15&13.6\\
0921880301&04:30:02.04&-61:22:25.5&2023-06-13&132.8\\
0921880401&04:30:52.45&-61:33:54.6&2023-06-15&132.9\\
0921880501&04:32:01.22&-61:17:33.4&2023-06-25&78.0\\
0921880601&04:32:32.01&-61:30:22.5&2023-06-27&80.0\\
0921880801&04:30:52.45&-61:33:54.6&2023-06-29&76.9\\
0921880101&04:32:01.22&-61:17:33.4&2023-10-17&117.9\\
0921880201&04:32:32.01&-61:30:22.5&2023-11-22&122.6\\
0921880701&04:30:02.04&-61:22:25.5&2023-11-24&85.9\\
 \hline
\end{tabular}
\end{table} 

\begin{figure} 
        \centering 
           \includegraphics[width=0.46\textwidth]{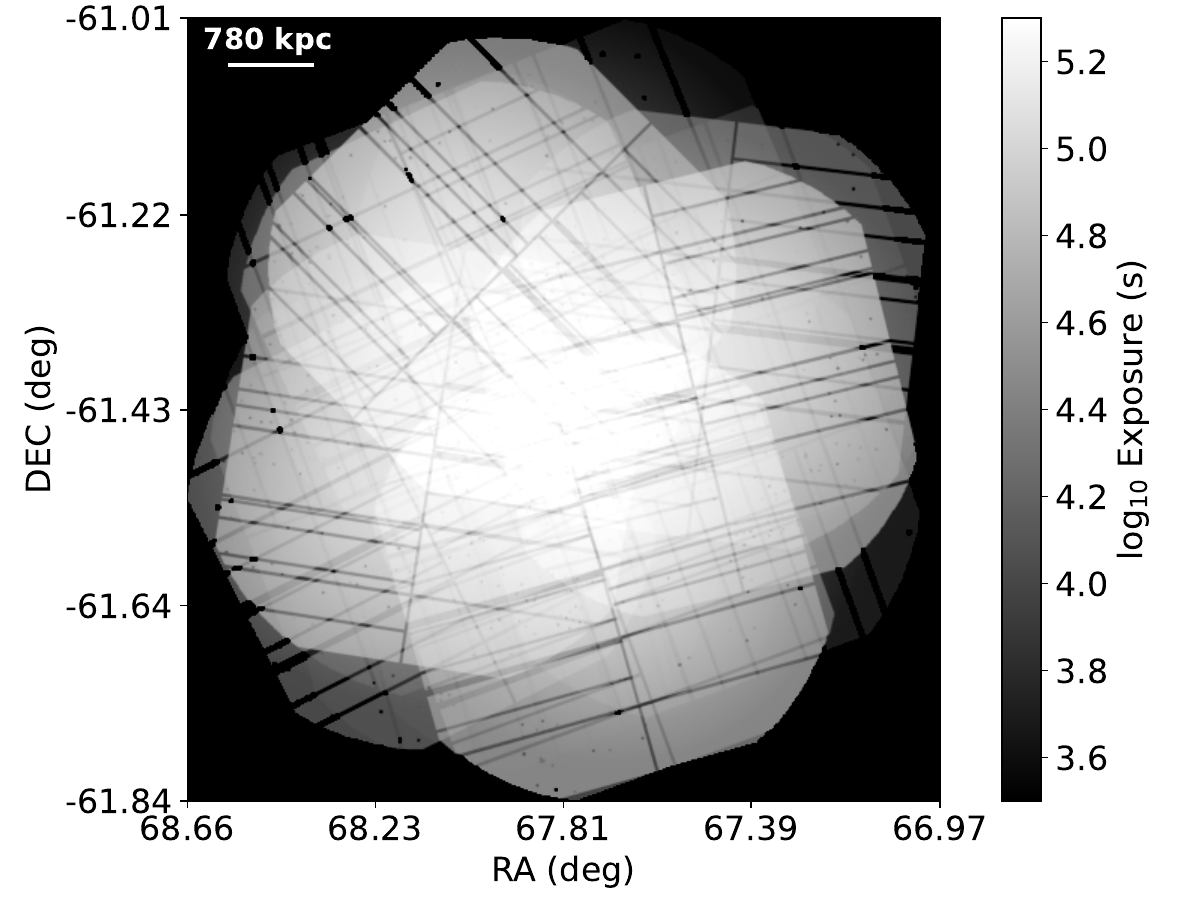}\\
           \includegraphics[width=0.46\textwidth]{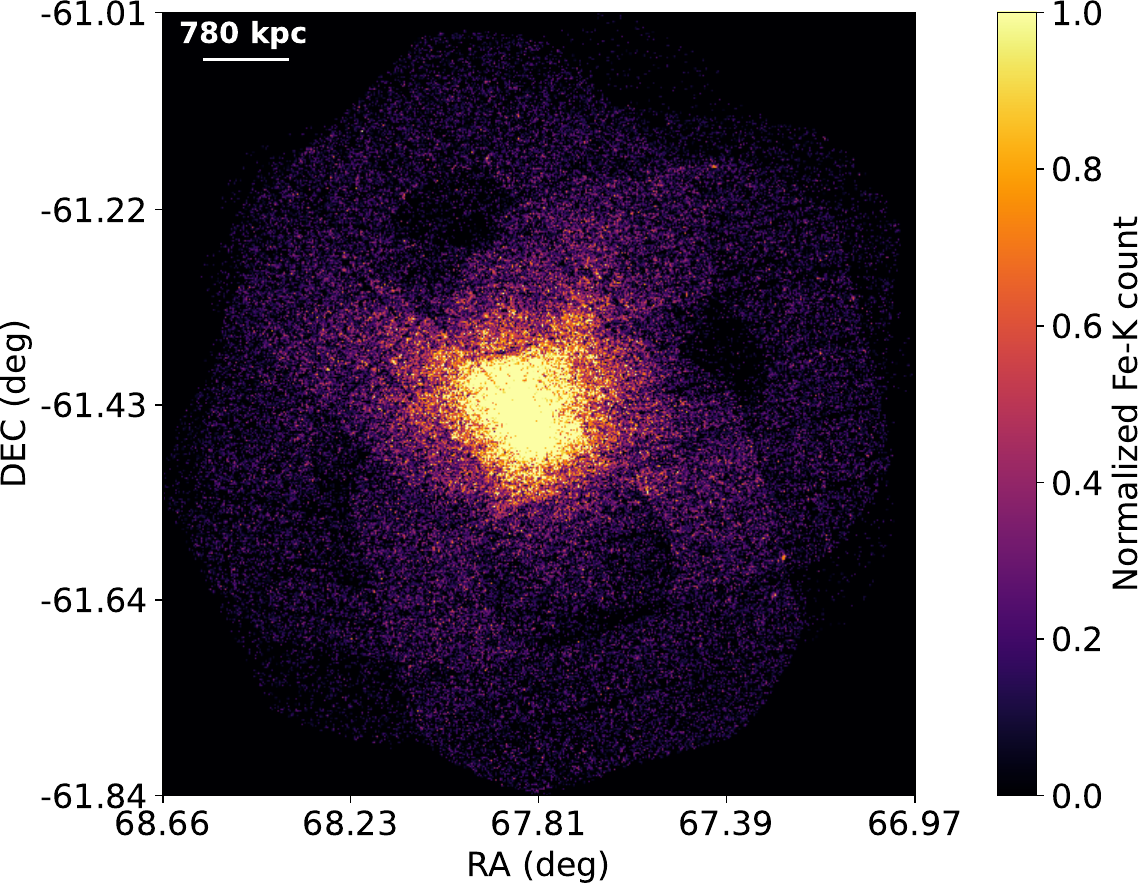}\\ 
           \includegraphics[width=0.46\textwidth]{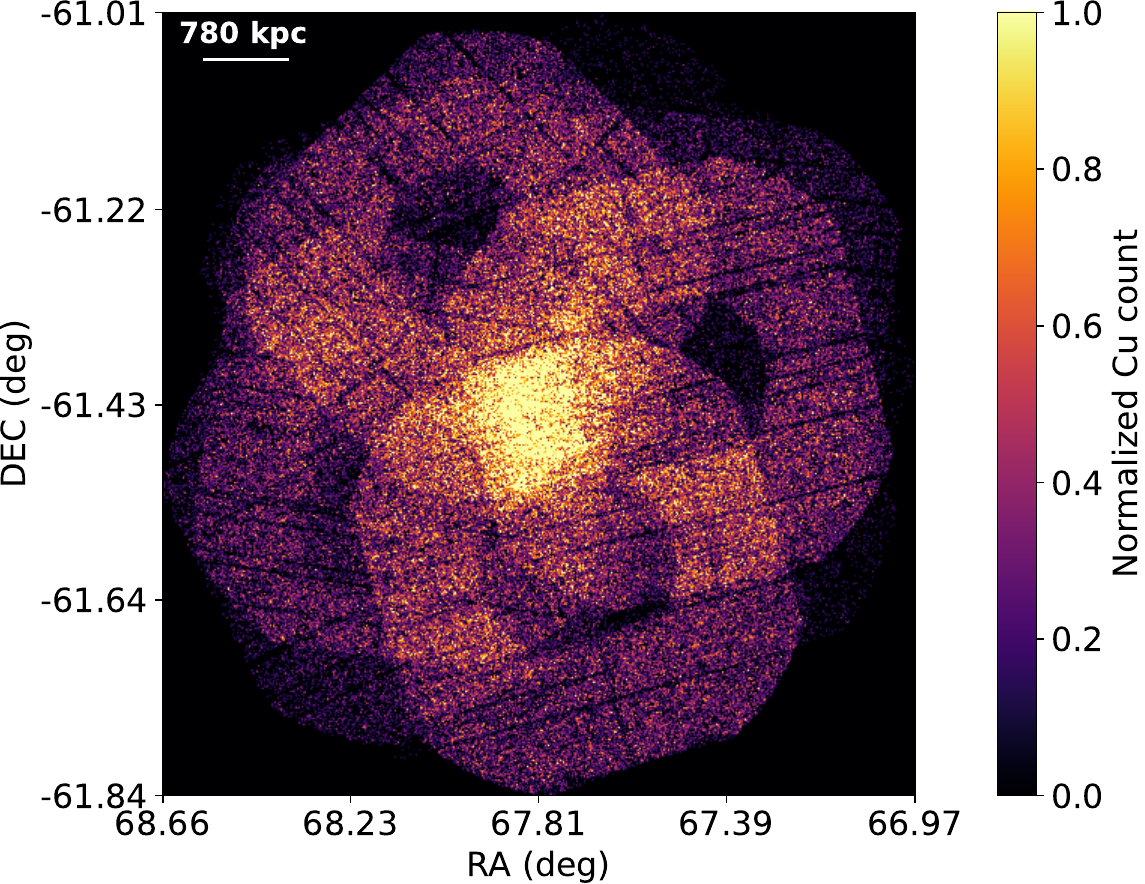} 
        \caption{\emph{Top panel:} total exposure time (s) in the 0.5-9.25 keV energy range. \emph{Middle panel:}  Normalized Fe-K count map, showing the number of counts in each 1.59 arcsec pixel in the Fe-K complex with the instrumental Cu ring included to indicate in which regions the redshift can be measured. \emph{Bottom panel:}  Normalized background Cu count map, showing the number of counts in each 1.59 arcsec pixel in the Cu~K$\alpha$ line energy range.} \label{fig_xray_maps2} 
\end{figure} 

The A3266 galaxy cluster is an excellent candidate for studying the ICM velocity distribution. 
It is one of the most brilliant clusters in the sky \citep{edg90} and one of the brightest merging clusters with temperatures exceeding 7 keV. 
ASCA observations of the cluster provide evidence of a merger, detecting a temperature variation along the merger axis \citep{hen00}. 
{\it Chandra} observations identified a cooler filamentary region centered on the central cD galaxy and aligned along the merger axis \citep{hen02}. 
\citet{fin06} estimated the mass of this cooler, denser material to be $1.3 \times 10^{13}$ M$_{\odot}$ with high metallicity using {\it XMM-Newton} observations, suggesting it results from a merger with a mass ratio of 1:10. 
\citet{sau05} proposed a similar scenario, attributing the low-entropy material to a merger in a direction close to the plane of the sky.

Numerical simulations computed by \citet{sau05} present two possible scenarios for the merger: the subcluster may have entered from the southwest, passing the core 0.8 Gyr ago and now nearing turnaround, or it could have merged from the northeast and is now exiting to the southwest after passing the core 0.15-0.20 Gyr ago. 
\citet{ett19} analyzed {\it XMM-Newton} data out to the cluster's virial radius, finding masses of $M_{500} = 8.8 \times 10^{14}$ M$_{\odot}$ and $M_{200} = 15 \times 10^{14}$ M$_{\odot}$, with radii of $R_{500} = 1.43$ Mpc and $R_{200} = 2.33$ Mpc, assuming an NFW mass model \citep{nav96}. 
\citet{deh17} measured spectroscopic redshifts of 1300 sources within the cluster, identifying that the cluster core, which can be split into two components, decomposes into six groups and filaments to the north. 
They concluded that a range of continuous dynamical interactions are taking place, rather than a simple NE-SW merger.

A3266 was the target of a calibration observation for the new eROSITA X-ray telescope \citep{pre21} during its calibration and performance verification (CalPV) phase \citep{den20}. 
In their data analysis, \citet{san22} confirmed that the source is not a simple merging cluster but rather involves three different systems merging with the main body: the NE structure, the NW structure, and the W structure. 
These systems are seen as low-entropy material associated with higher-metallicity gas. 
However, the data did not conclusively reveal whether there is a merger shock on the west side of the central core. 
Pressure jumps in the southeast and west directions were identified towards the outskirts of the cluster. 
The presence of several subclusters that are merging or will soon merge makes A3266 an ideal laboratory for detailed study of the ICM velocity structure within such a system.

We present an analysis of the ICM velocity structure within the A3266 galaxy cluster using {\it XMM–Newton} observations. 
The outline of this paper is as follows. 
We describe the data reduction process in Section~\ref{sec_dat} and analyze the X-ray images in Section~\ref{sec_ima}.
The fitting procedure and the results are shown in Section \ref{sec_fits} while a discussion of the results is included in Section~\ref{sec_dis}. 
Finally, Section~\ref{sec_con} presents the conclusions and summary. 
Throughout this paper we assume the distance of A3266 to be $z=0.0594$ \citep{deh17} and a concordance $\Lambda$CDM cosmology with $\Omega_m = 0.3$, $\Omega_\Lambda = 0.7$, and $H_{0} = 70 \textrm{ km s}^{-1}\ \textrm{Mpc}^{-1} $.

\section{A3266 XMM-Newton observations}\label{sec_dat}

\subsection{EPIC-pn spectra}
\begin{figure} 
        \centering 
           \includegraphics[width=0.47\textwidth]{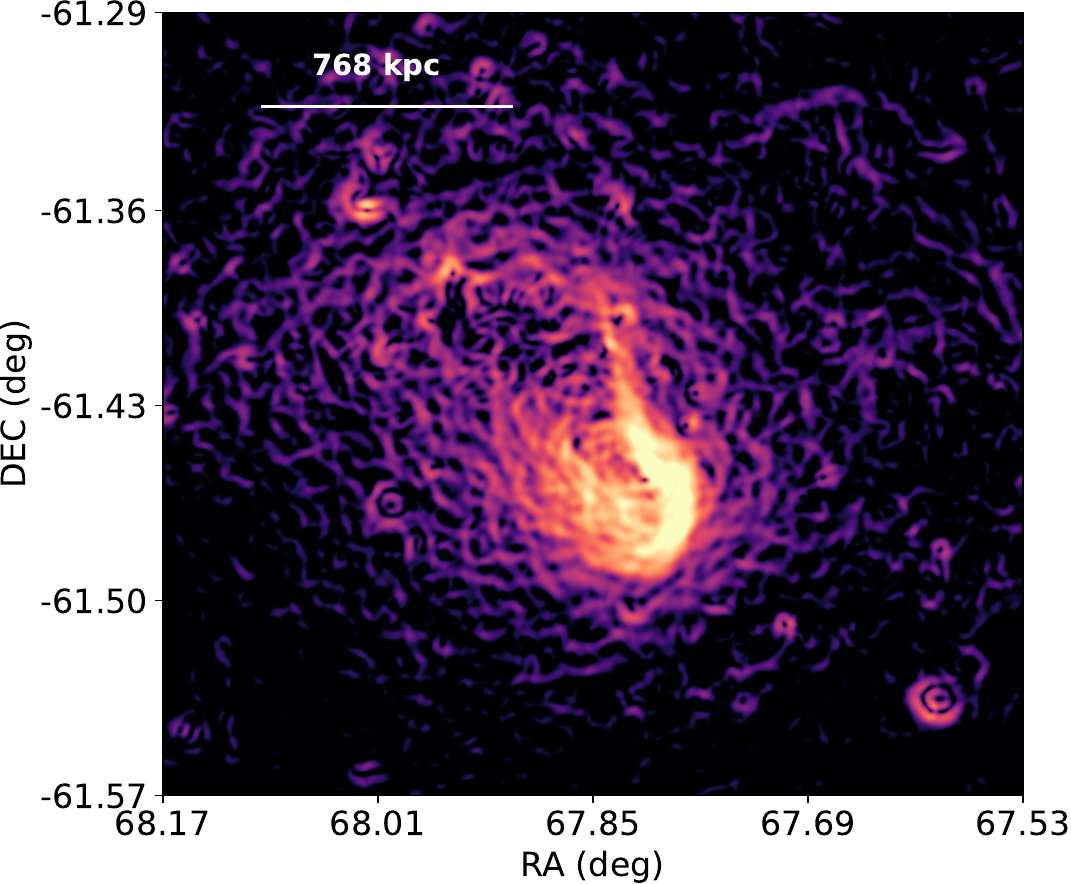}\\  
           \includegraphics[width=0.47\textwidth]{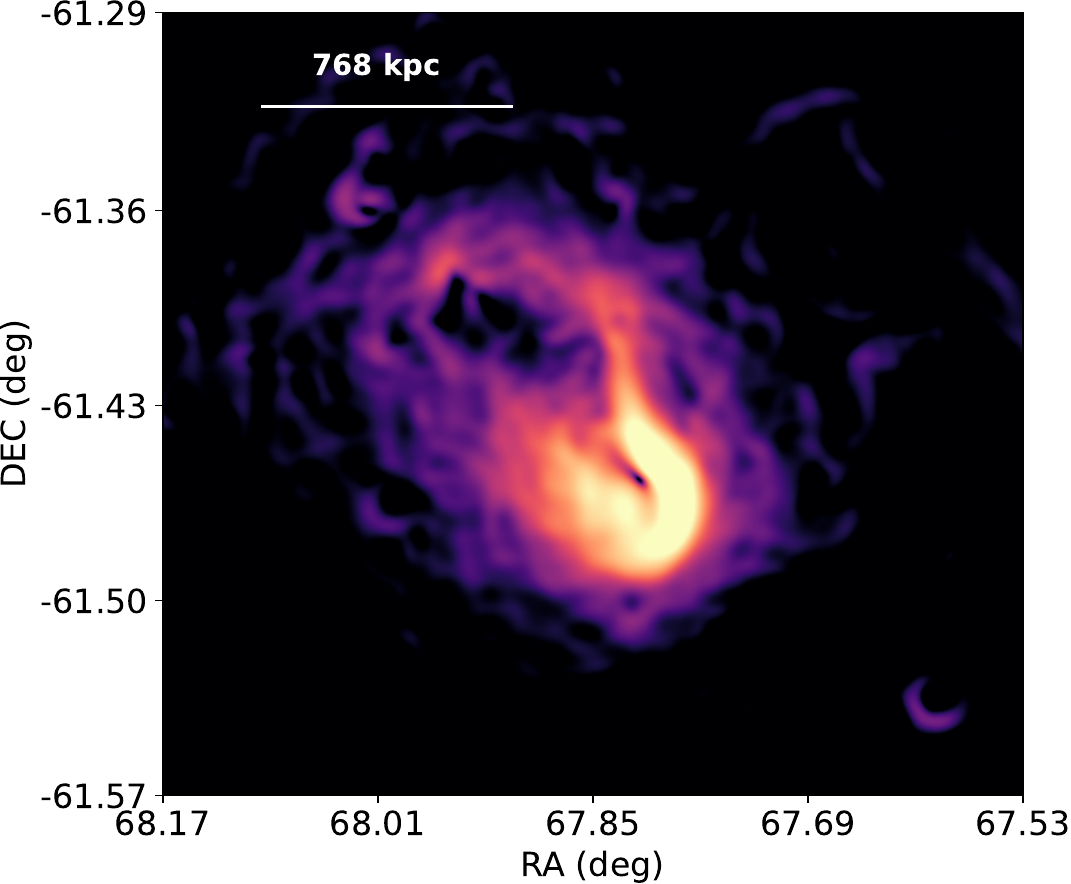} 
        \caption{Edge-filtered X-ray images using a Gaussian Gradient Magnitude (GGM) filter on scales of 4 pixels (\emph{top panel}) and 8 pixels (\emph{bottom panel}). }\label{fig_xray_maps3} 
\end{figure}

We analyzed eight A3266 observations taken between June and November 2023 (PI: E. Gatuzz, ID:092188). 
Additionally, we included ten observations that are available in the archive. 
Table~\ref{tab_obsids} lists information on the observations, including IDs, coordinates, dates, and clean exposure times. 
All {\it XMM-Newton} European Photon Imaging Camera \citep[EPIC-pn,][]{str01} spectra were reduced with the Science Analysis System (SAS\footnote{\url{https://www.cosmos.esa.int/web/xmm-newton/sas}}, version 19.1.0). 
We processed the observations with the {\tt epchain} SAS tool, including only single-pixel events (PATTERN==0) and filtering the data with FLAG==0 to avoid bad pixels and regions close to CCD edges. 
A 1.0 cts/s rate threshold was applied to filter out bad time intervals from flares.

To measure the velocity structure within the A3266 cluster, we applied the technique described in \citet{san20} and \citet{gat22a} to calibrate the absolute energy scale of the EPIC-pn detector using the instrumental background X-ray lines identified in the detector spectra. 
This calibration method includes corrections for (1) the average gain of the detector during the observation, (2) the spatial gain variation across the detector over time, and (3) the energy scale as a function of detector position and time. 
We created X-ray images and identified point sources using the SAS task {\tt edetect\_chain}, with a likelihood parameter {\tt det\_ml} $> 10$. 
In the following analysis we have excluded the point sources.

\begin{figure}
        \centering 
           \includegraphics[width=0.48\textwidth]{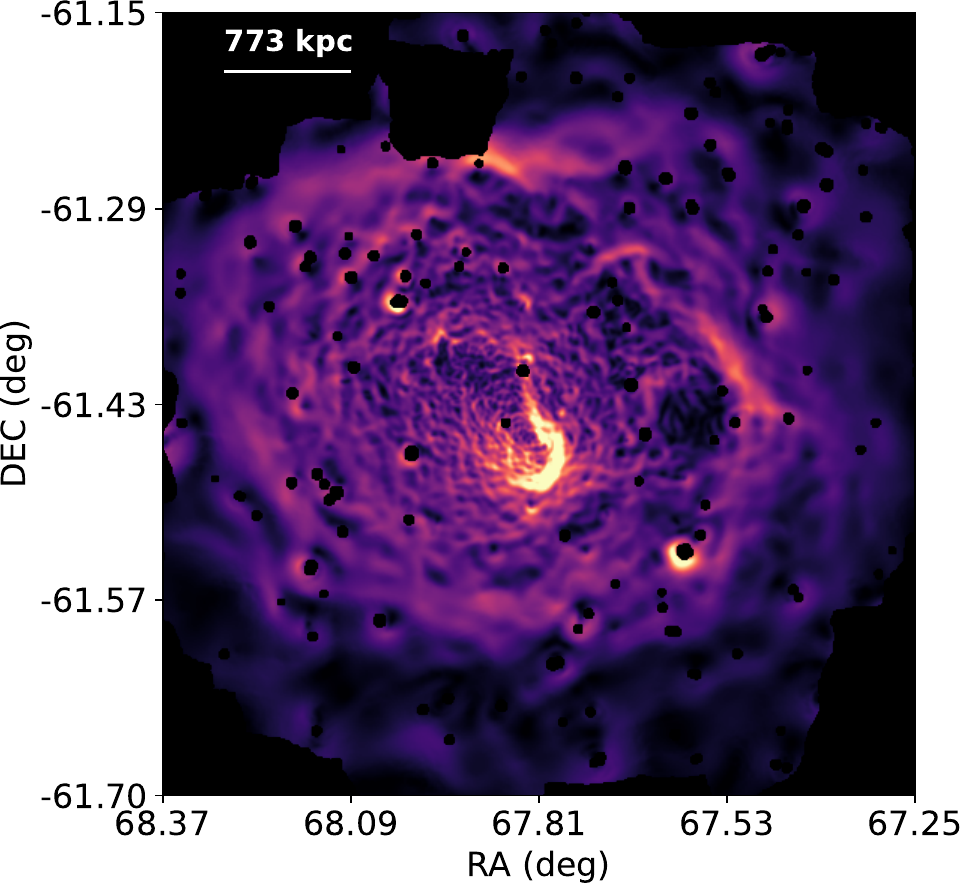}
        \caption{Edge-filtered X-ray images using an adaptive Gaussian Gradient Magnitude (GGM) filter from \citet{san22a} using a signal to noise ratio of 32, where the scale shows the gradient as the $\log(10)$ change per 4 arcsec pixel.  } \label{fig_xray_maps4} 
\end{figure}

\section{X-ray images}\label{sec_ima} 
The top panel of Figure~\ref{fig_xray_maps1} shows an RGB image of the cluster, with energy bands corresponding to 0.2-0.5 keV (R), 0.5-1.0 keV (G), and 1.0-2.0 keV (B). 
The exposure-corrected images were smoothed with a Gaussian of $\sigma=1$. 
The bottom panel of Figure~\ref{fig_xray_maps1} displays the X-ray exposure-corrected image in the 0.5-9.25 keV energy band. 
A Gaussian smoothing of $\sigma=4$ pixels was applied, and the 10-level contour of the X-ray image is shown in white.

The top panel of Figure~\ref{fig_xray_maps2} shows the total exposure time in the 4.0-9.25 keV energy range. 
The middle panel displays the normalized number of counts in each 1.59 arcsec pixel within the Fe-K complex (6.50-6.90 keV), after subtracting the neighboring continuum images (6.06-6.43 keV and 6.94-7.16 keV, rest frame). 
The instrumental Cu ring is included to indicate in which regions the redshift can be measured.
This image demonstrates a good distribution of counts in the cluster center region, allowing for detailed analysis of the Fe-K complex. 
The bottom panel shows the normalized number of counts in each 1.59 arcsec pixel around the Cu background line. 
The plot reveals the Cu hole described in \citet{san20} for each exposure. 
Since velocity measurements can only be obtained using the Cu background line, it is crucial to have a large number of counts in regions where the Fe-K complex redshift will be measured.

\begin{figure} 
\centering 
\includegraphics[width=0.485\textwidth]{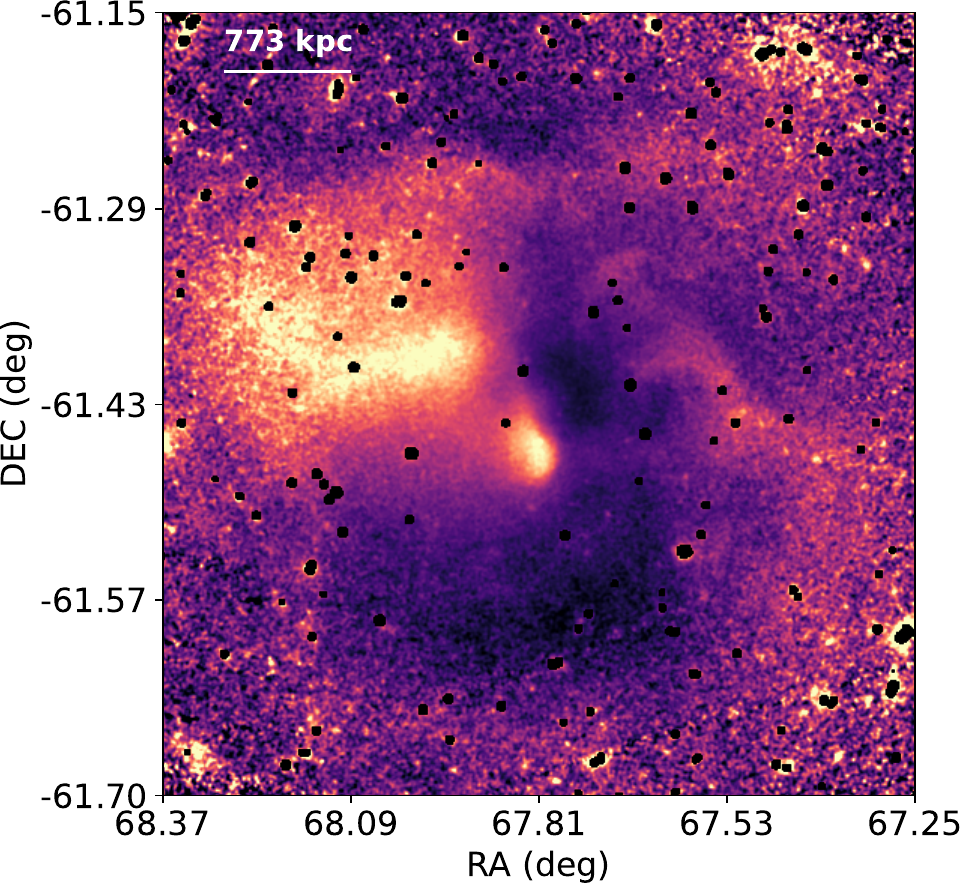}
\caption{Subtracted fractional difference from the average at each radius for the 0.5-2~keV X-ray surface brightness. The point-like sources were excluded in the analysis. The X-ray image was smoothed by a 2$\sigma$ Gaussian.} \label{fig_surf_bri} 
\end{figure}

To study the structures within the galaxy cluster in detail, we filtered the data using a Gaussian Gradient Magnitude (GGM) filter \citep[GGM,][]{san16a,san16b}. 
Figure~\ref{fig_xray_maps3} shows the gradient magnitude of the X-ray image convolved with Gaussians of scales 4 pixels (8 arcsec, top panel) and 8 pixels (16 arcsec, bottom panel). 
Point sources were cosmetically filled using the values of neighboring pixels before filtering. 
A 700-800 kpc elliptical surface brightness edge surrounding the core is clearly shown. 
The image does not accurately depict the relative magnitudes of features but highlights these structures.

We also applied the adaptively smoothed GGM filter developed in \citet{san22}. 
To apply this filter, the image and exposure map were convolved by a Gaussian with a $\sigma$ corresponding to a radius containing 1024 counts. 
The $\log_{10}$ value of this map was then taken, and the gradient was calculated by taking the difference between neighboring pixels along the two axes and adding them in quadrature. 
Figure~\ref{fig_xray_maps4} shows the resulting image, clearly revealing the sharp edge near the cluster core.

Figure~\ref{fig_surf_bri} shows the fractional difference in the 0.5 to 2 keV surface brightness from the average at each radius. 
The black regions correspond to point-like sources whose sizes have been artificially increased due to image smoothing. 
The X-ray image was smoothed by a 2$\sigma$ Gaussian.
Multiple clear sharp discontinuities are visible, which have also been identified by \citet{san22}. 
We analyze these structures in detail in subsection~\ref{sub_sur_fits}. 


\section{Spectral fitting procedure}\label{sec_fits}

The spectra were fitted in the 4.0-9.25 keV energy band. For each spatial region analyzed, we combined the spectra from different observations, using the number of counts in the fitting spectral region as the weighting factor. 
We modeled the ICM X-ray emission with the {\tt apec} thermal emission model version 3.0.9 \citep{fos19}, which includes corrections derived from the {\it Hitomi} observations \citep{hit18}. 
A {\tt tbabs} component \citep{wil00} was also included to account for the Galactic X-ray absorption, corresponding to $2.26\times 10^{20}$ cm$^{-2}$ \citep{kal05}. 
However, we note that for the energy range analyzed, the absorbing component has no effect on the modeling \citep{gat22a,gat22b,gat23a}. 
The free parameters in the model are the temperature, metallicity (i.e., metal abundances with He fixed at cosmic levels), redshift, and normalization. 

For the X-ray spectral fits, we used the {\it xspec} data fitting package (version 12.14.0\footnote{\url{https://heasarc.gsfc.nasa.gov/xanadu/xspec/}}). We assumed {\tt cash} statistics \citep{cas79}. Errors are quoted at the 1-$\sigma$ confidence level unless otherwise stated. Abundances are given relative to \citet{lod09}. 
As background components, we included Cu-$K\alpha$, Ni-$K\alpha$, Zn-$K\alpha$, and Cu-$K\beta$ instrumental emission lines, as well as a power-law component with its photon index fixed at 0.136 \citep[the average value obtained from the archival observations analyzed in][]{san20}.

\subsection{Spectral maps}\label{spec_fit_sec}

\begin{figure*}
\centering 
\includegraphics[width=0.485\textwidth]{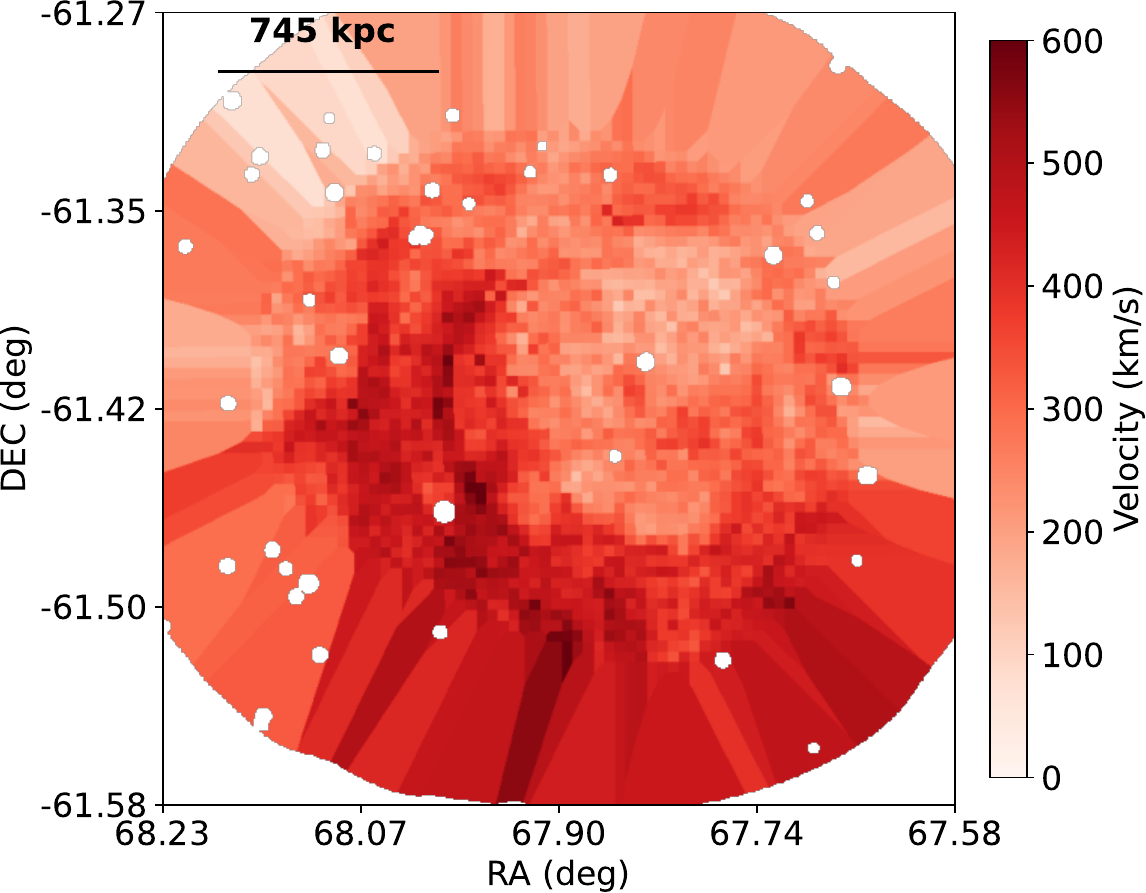}
\includegraphics[width=0.485\textwidth]{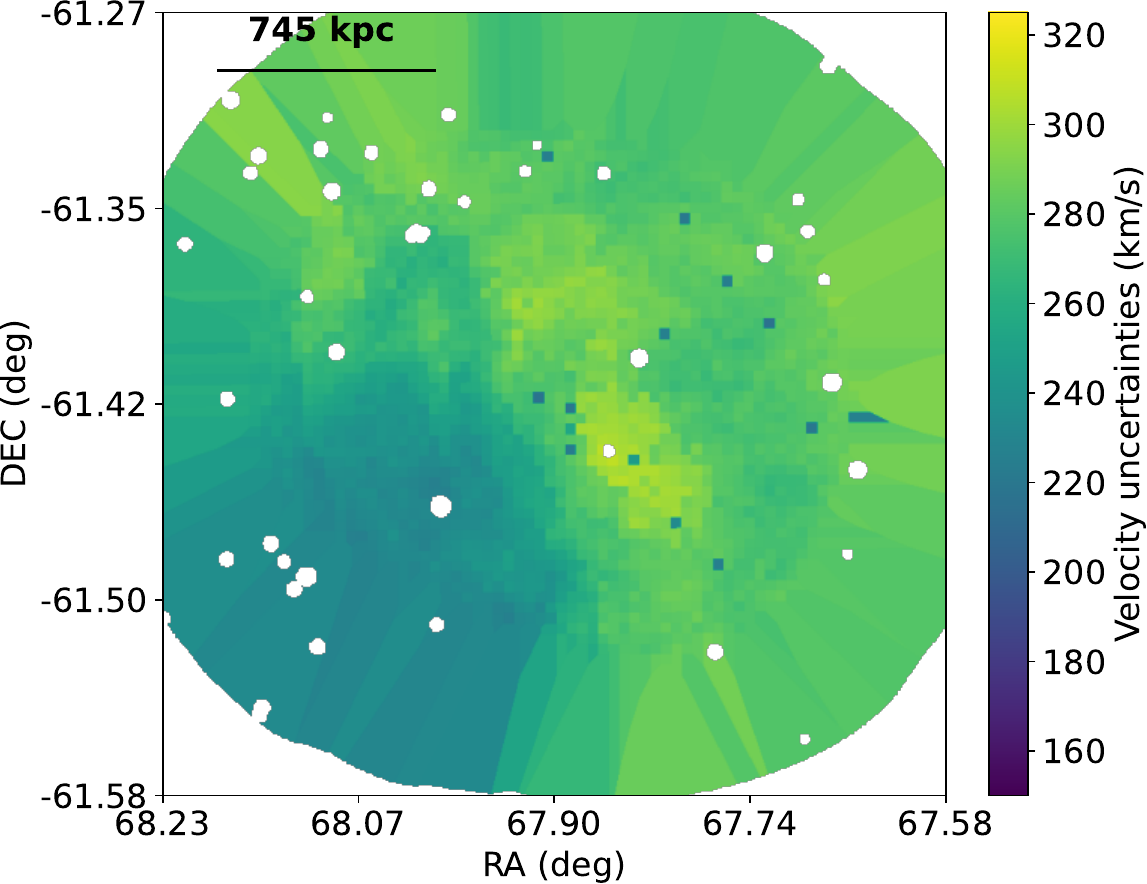}
\caption{\emph{Left panel:} velocity map (km/s) relative to the A3266 cluster ($z=0.0594$). Maps were created by moving 2:1 elliptical regions (i.e., rotated to lie tangentially to the nucleus) containing $\sim$ 750 counts in the Fe-K region. \emph{Right panel:} 1$\sigma$ velocity uncertainties. White circles correspond to point sources which were excluded from the analysis. The spectral fitting is done in the $4-9.25$~keV energy range.} \label{fig_spec_map1} 
\end{figure*}

\begin{figure}
\centering 
\includegraphics[width=0.485\textwidth]{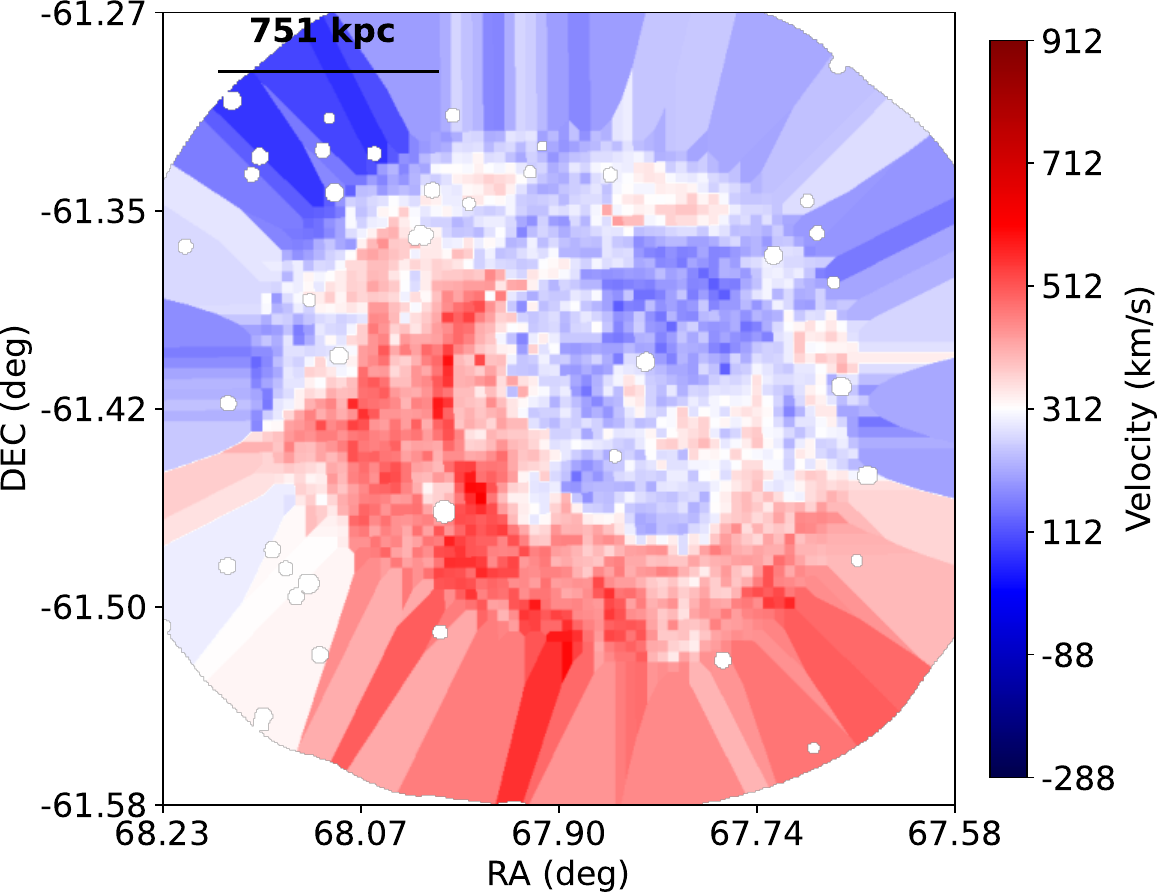}
\caption{Velocity map (km/s) relative to the A3266 cluster, after subtracting the mean value $312$~km/s obtained from the spectral map described in Section~\ref{fig_spec_map1}. } \label{fig_spec_rest} 
\end{figure}

\begin{figure} 
\centering 
\includegraphics[width=0.485\textwidth]{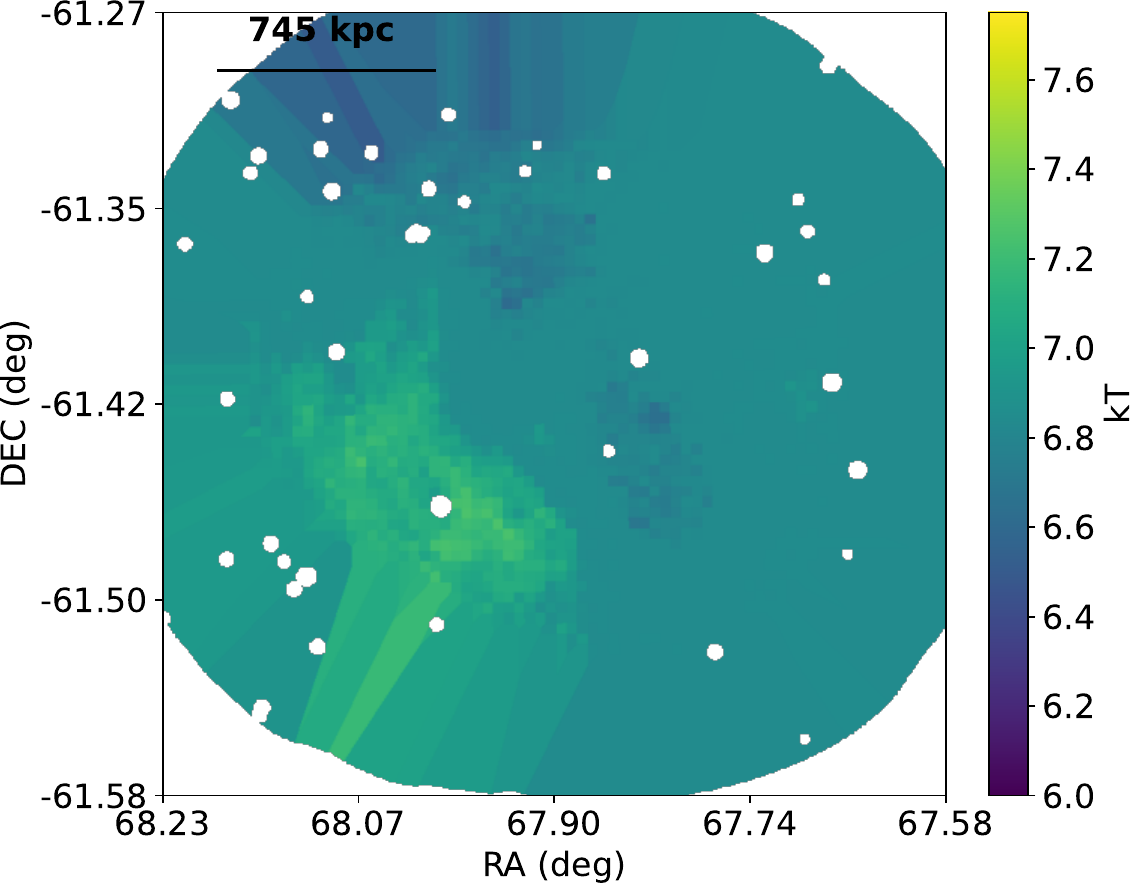}\\
\includegraphics[width=0.485\textwidth]{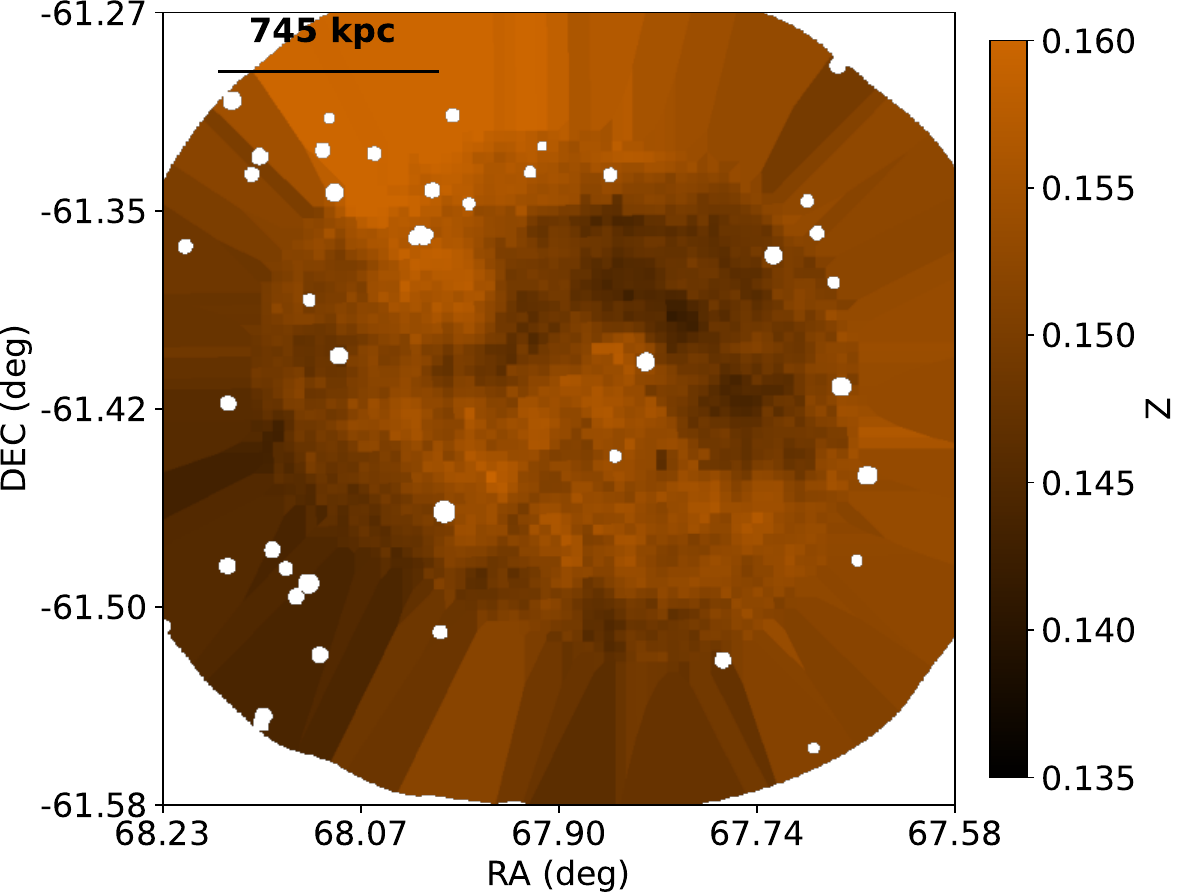}\\
\includegraphics[width=0.485\textwidth]{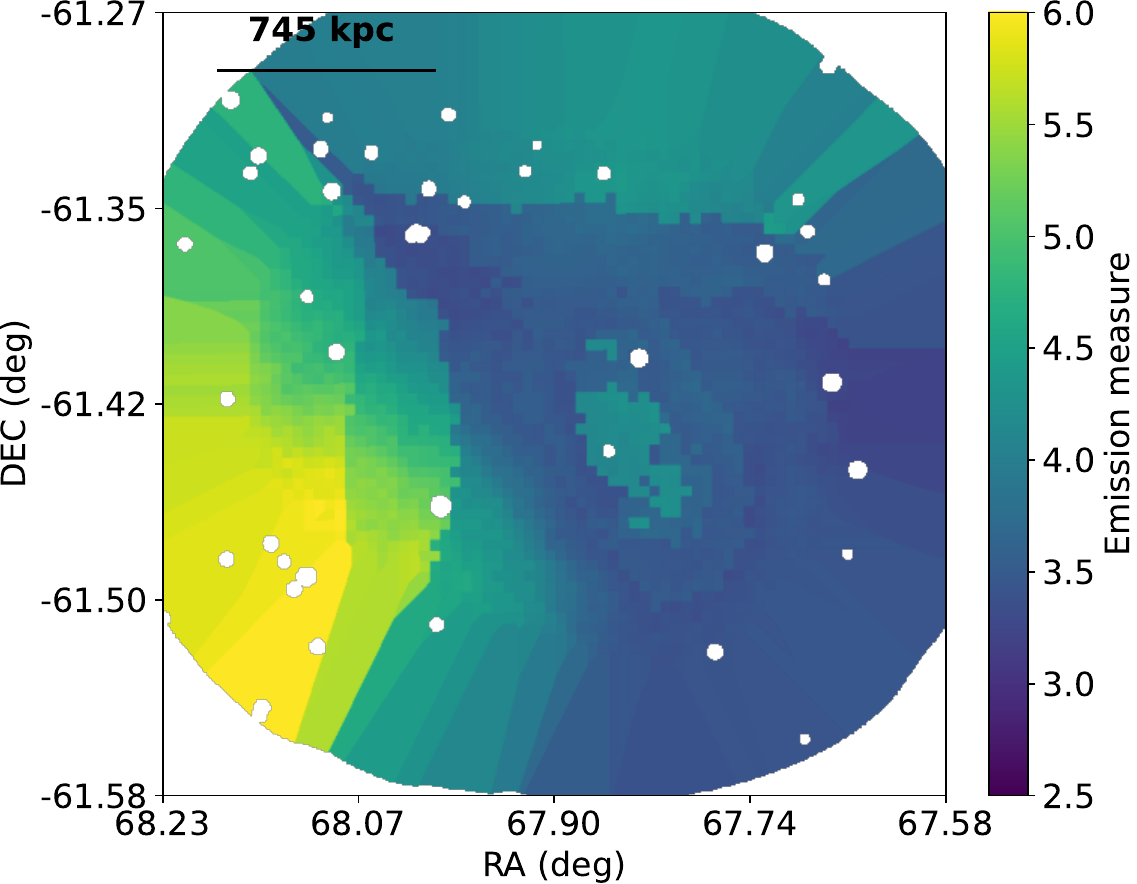}
\caption{\emph{Top panel:} temperature map in units of keV. \emph{Middle panel:} Abundance map relative to abundances from \citet{lod09}. \emph{Bottom panel:} emission measure map. The spectral fitting is done in the $4-9.25$~keV energy range.} \label{fig_spec_map2} 
\end{figure}

We created maps of the cluster spectral properties using a method developed in \citet{san20}, based on fitting spectra from dynamically sized ellipses. 
To create these maps, we used elliptical regions with a 2:1 axis ratio, rotated so that the longest axis lies tangentially to the vector pointing to the central core. 
The radii of the ellipses adaptively change in a grid with a spacing of 0.25 arcmin to achieve approximately 1000 counts in the Fe-K complex after continuum subtraction, thereby reducing the velocity uncertainties.
The semi-major axis range from $6.1$~arcmin to $2.0$~arcmin.
We combined the spectra obtained per region for all observations and created weighted-average ancillary response files (ARFs) while using the same response matrix for all the pixels. 
The spectra were then fitted as described in Section~\ref{sec_fits}.

Figure~\ref{fig_spec_map1} shows the velocity map (left panel) and the velocity uncertainties (right panel) obtained. 
We use a blue-red color scale for the velocity map to indicate whether the gas is moving towards the observer (blue) or away (red) to improve clarity.  
We found a mean velocity of $\sim 312$~km/s for the entire map.
For illustrative purposes, Figure~\ref{fig_spec_rest} shows the velocity map relative to the A3266 cluster, after subtracting the mean value $312$~km/s obtained from the spectral map.
The mean optical velocity for galaxies from \citet{deh17} for all regions within the entire map is $\sim 1727.5$~km/s, larger than the ICM overall velocity. 
Figure~\ref{fig_spec_map2} shows the spectral maps for temperature (top panel), metallicity (middle panel), and emission measure (bottom panel). 
We found that the metallicity is not uniform within the cluster, decreasing towards the northwest, corresponding to the low fractional difference shown in Figure~\ref{fig_surf_bri}. 
The metallicities we found are lower than those measured by \citet{san22}. 
However, it is important to note that our analysis only includes the high-energy band ($>4$~keV), so the absolute abundances are not entirely reliable. 
The temperature map tends to be more uniform, but as indicated before, the actual temperature distribution cannot be accurately determined since we are modeling only the high-energy band. 

\subsection{Velocity radial profiles}\label{fit_rings}

\begin{table}
\scriptsize 
\caption{\label{tab_circular_fits}A3266 cluster best-fit parameters for annular regions. }
\centering
\begin{tabular}{ccccccc}
\\
Region &\multicolumn{5}{c}{{\tt apec} model}  \\
\hline
 &$kT$ & Z& $z$   & $norm$   & cstat/dof\\ 
  &   & &  ($\times 10^{-3}$) &   ($\times 10^{-3}$)  \\  
1&$7.19_{-0.45}^{+0.64}$&$0.19\pm 0.03$&$12.69_{-0.45}^{+0.41}$&$59.94_{-2.25}^{+2.36}$&$1470/1199$\\
2&$7.00_{-0.26}^{+0.48}$&$0.21\pm 0.02$&$17.74_{-0.49}^{+0.37}$&$61.01\pm 1.63$&$1498/1199$\\
3&$7.46\pm 0.44$&$0.22\pm 0.02$&$19.30\pm 0.49$&$61.65\pm 1.57$&$1494/1199$\\
4&$7.45\pm 0.46$&$0.21\pm 0.02$&$19.80\pm 0.52$&$60.22\pm 1.48$&$1484/1199$\\
5&$7.34\pm 0.49$&$0.18\pm 0.02$&$19.60\pm 0.55$&$62.25\pm 1.90$&$1529/1199$\\
6&$7.34\pm 0.51$&$0.20\pm 0.02$&$18.38\pm 0.54$&$59.98_{-1.81}^{+1.88}$&$1490/1199$\\
7&$7.40\pm 0.51$&$0.19\pm 0.02$&$19.07_{-0.54}^{+0.57}$&$58.43_{-1.84}^{+1.93}$&$1499/1199$\\
8&$7.65\pm 0.33$&$0.21\pm 0.02$&$34.37\pm 0.63$&$61.13\pm 1.27$&$1405/1199$\\ 
\\ 
 \hline
\end{tabular}
\end{table}

   \begin{figure}
   \centering
\includegraphics[width=0.46\textwidth]{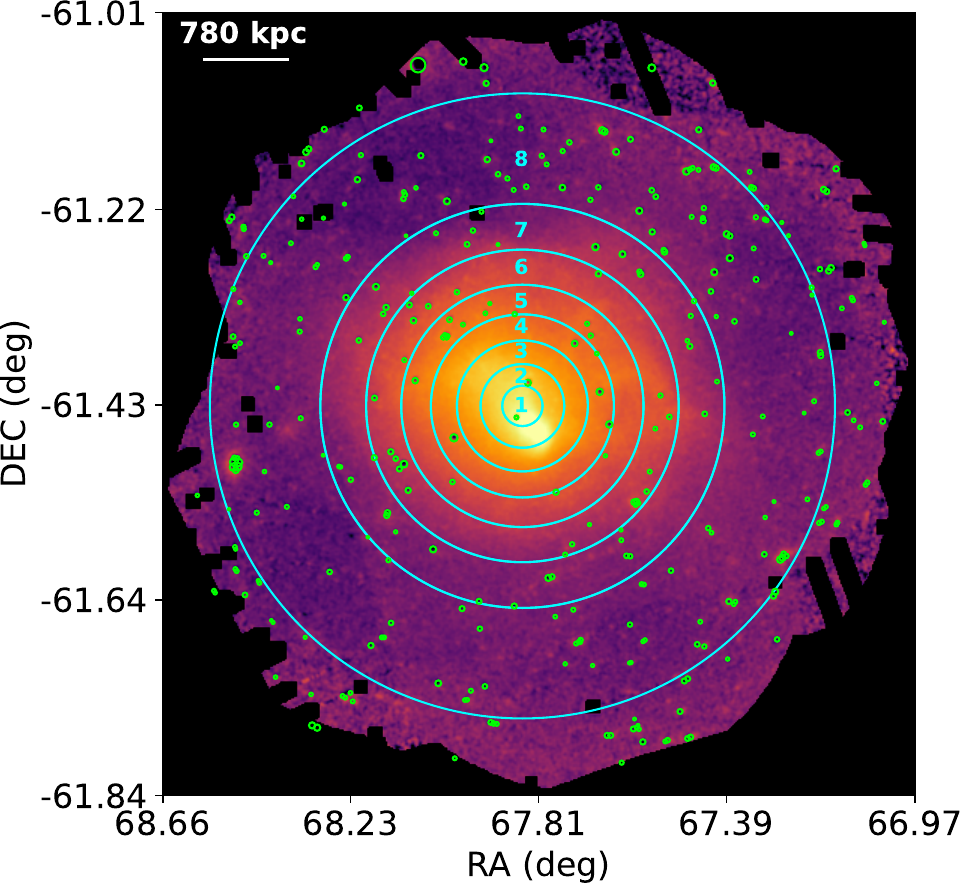}
\caption{A3266 annular regions analyzed in subsection~\ref{fit_rings}. Green circles correspond to point-like sources excluded from the analysis. } \label{fig_cas1_region} 
    \end{figure}

\begin{figure} 
\centering 
\includegraphics[width=0.46\textwidth]{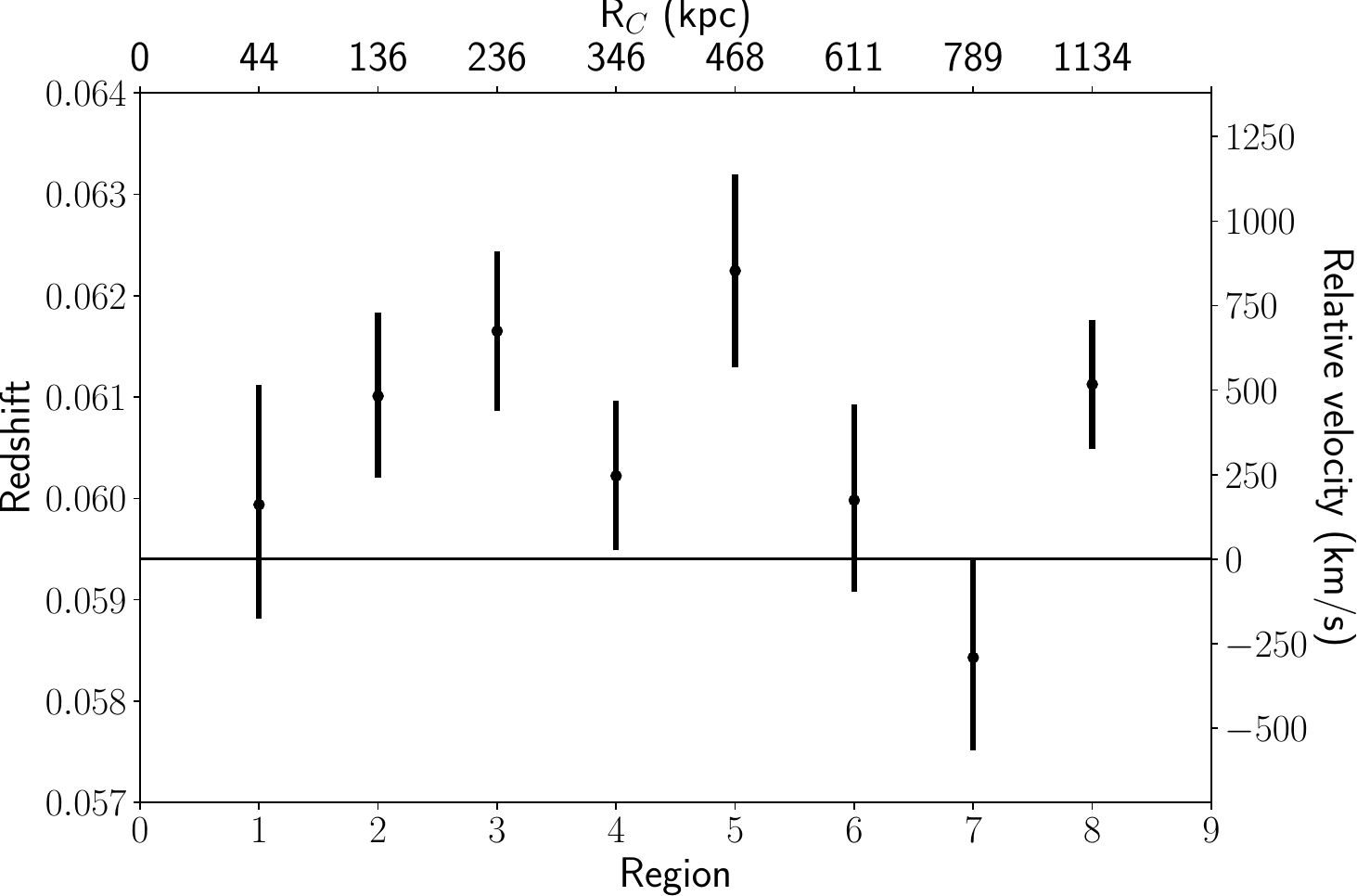} 
        \caption{Velocities obtained for each annular region described in subsection~\ref{fit_rings} (numbered from the center to the outside). The A3266 redshift is indicated with a horizontal line. The spectral fitting is done in the $4-9.25$~keV energy range.} \label{fig_cas1_resultsa} 
\end{figure}

\begin{figure} 
\centering  
\includegraphics[width=0.46\textwidth]{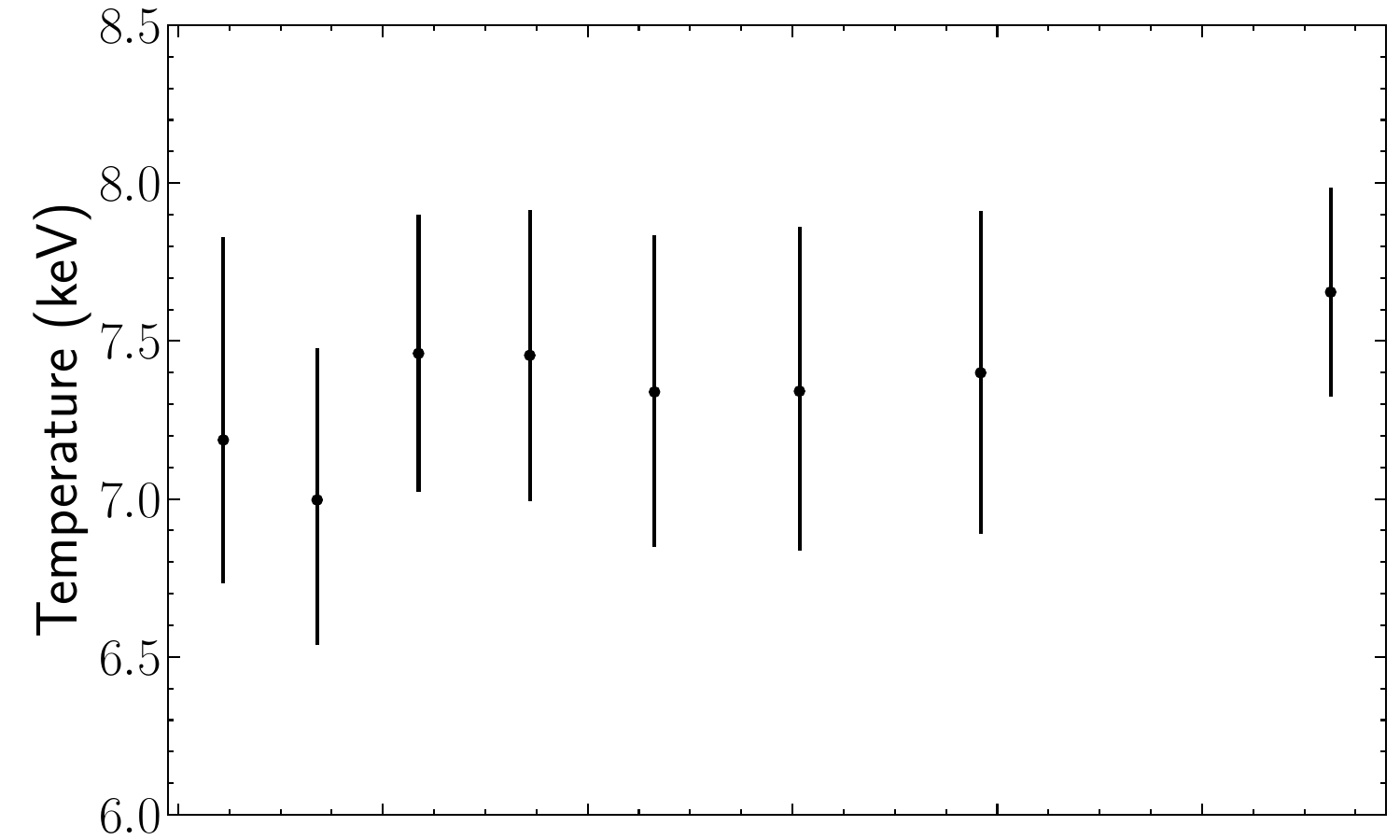}\\
\includegraphics[width=0.46\textwidth]{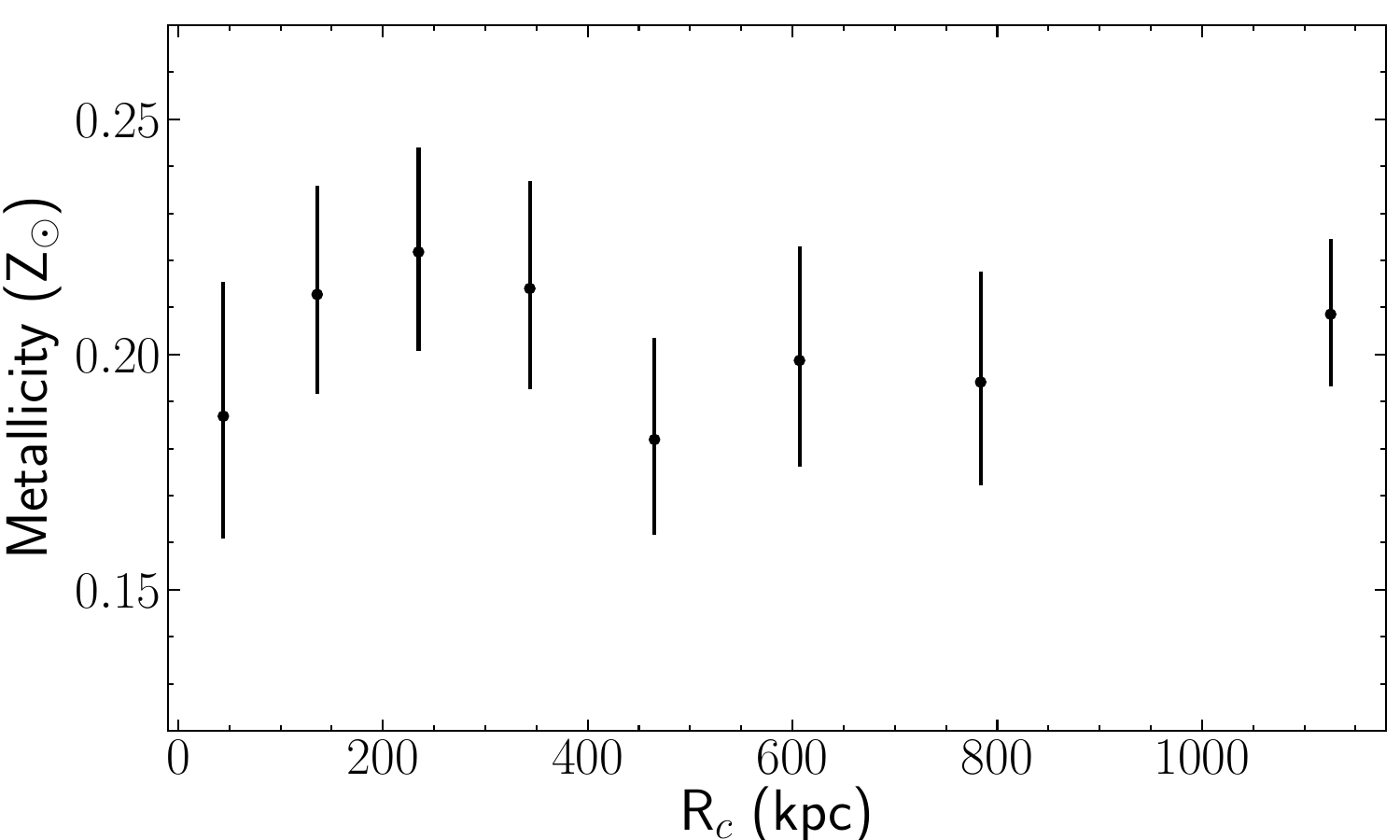} 
        \caption{Temperature (top panel) and metallicity (bottom panel) profiles obtained from the best fit results for each annular region described in subsection~\ref{fit_rings}. The spectral fitting is done in the $4-9.25$~keV energy range.} \label{fig_cas1_resultsb} 
\end{figure}

We proceeded to analyze annular regions to study the radial velocity structure. 
The thickness of these rings increases as the square root of the distance from the center of A3266. Figure~\ref{fig_cas1_region} shows the exact spatial regions analyzed. 
Table~\ref{tab_circular_fits} presents the best-fit results obtained for each region, numbered from the center outward, while Figure~\ref{fig_cas1_resultsa} displays the resulting velocity structure. 
We achieved velocity measurements with an accuracy of up to \(\Delta v \sim 190\) km/s (for ring 4). 
The largest redshift/blueshift relative to A3266 is $(805 \pm 284)$ km/s for ring 5. 
As observed in the spectral map (see subsection~\ref{spec_fit_sec}), we obtained a redshifted velocity distribution for almost all regions. 
For distances less than 240 kpc, the velocities tend to increase with radius. 
We found a mean velocity of $\sim 283$~km/s for all the regions.
We computed a $\chi^{2}$ test based on the null hypothesis that the velocities are consistent with a constant velocity model and obtained a $p-value=0.33$, thus the observed variations in velocity are not due to random chance.
Figure~\ref{fig_cas1_resultsb} shows the temperature (top panel) and metallicity (bottom panel) distribution. 
We found a uniform distribution for both parameters, similar to previous profiles computed by \citet{ghi19,san22a}. 
However, since our spectral analysis does not include the soft energy band, these values could be underestimated. 
Furthermore, the apparent correlation between the emission measure and metallicity is most likely an artifact as the lower the metallicity the larger should be the emission measure to obtain lines of comparable fluxes.

\subsection{Discontinuities in the surface-brightness}\label{sub_sur_fits}

\begin{figure}
   \centering
\includegraphics[width=0.46\textwidth]{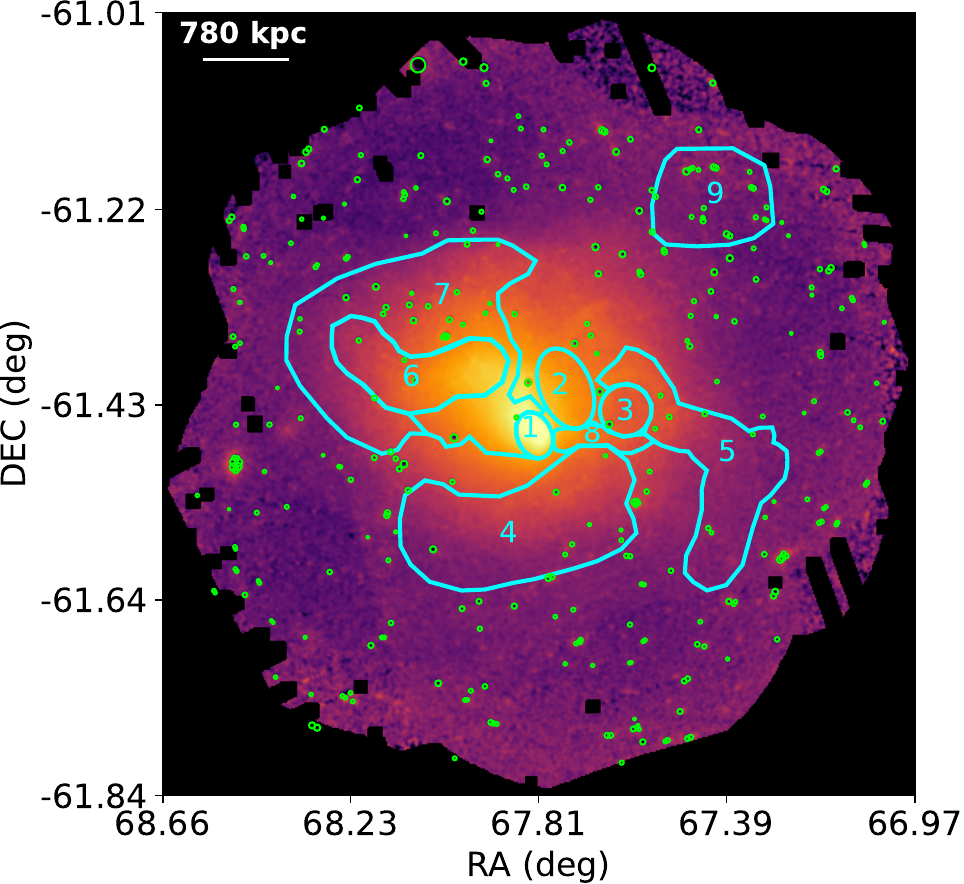}\\
\includegraphics[width=0.46\textwidth]{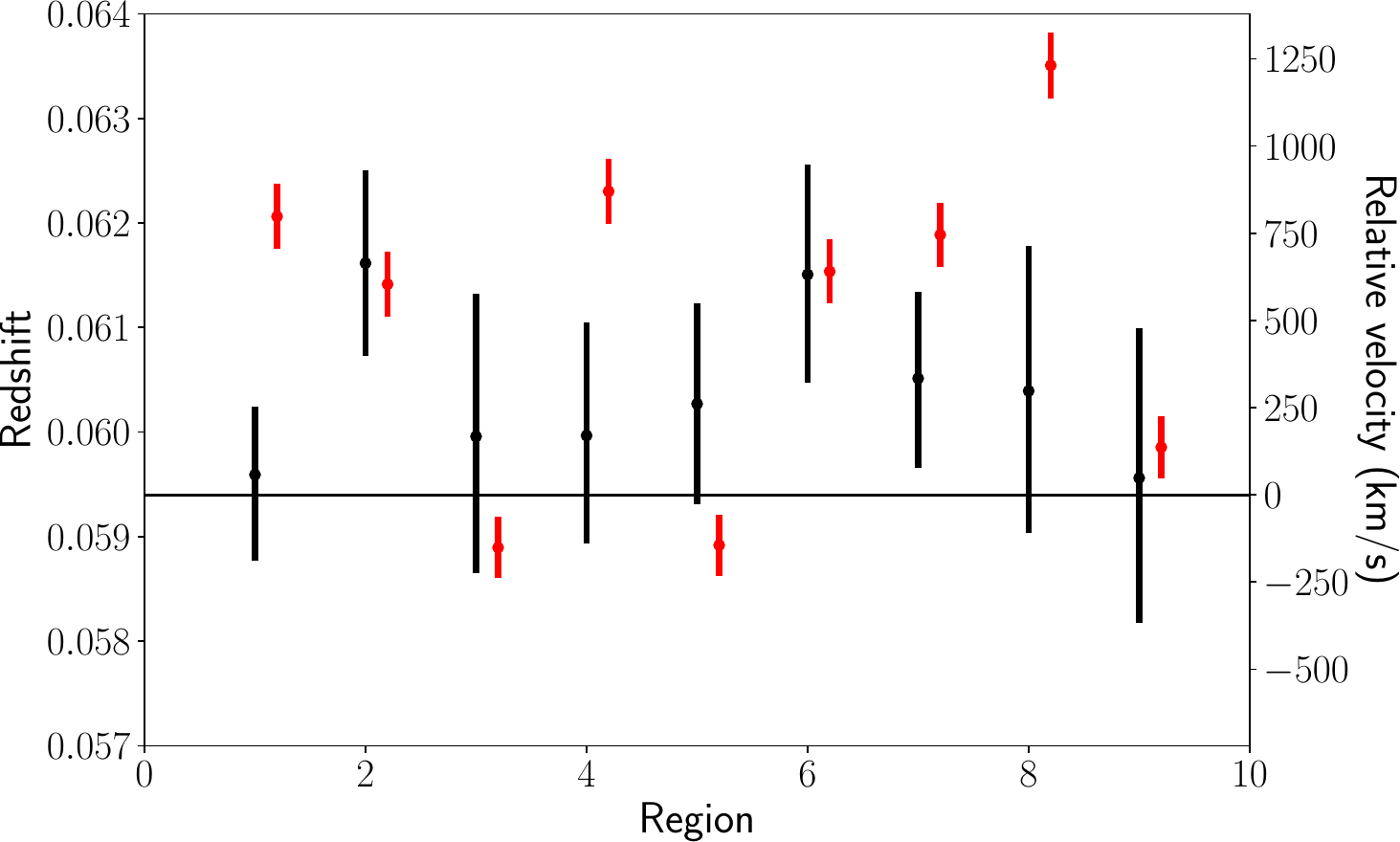} 
\caption{\emph{Top panel:} Manually selected regions following the fractional difference in 0.5-2~keV X-ray surface brightness. Green circles correspond to point-like sources excluded from the analysis. \emph{Bottom panel:} Velocities obtained for manually selected regions following the A3266 merging structure. The A3266 redshift is indicated with a horizontal line. The spectral fitting is done in the $4-9.25$~keV energy range. We have also included the mean optical redshift for galaxies from \citet{deh17} within each one of the regions obtained.} \label{fig_fit_sur1} 
    \end{figure}

\begin{figure} 
\centering  
\includegraphics[width=0.46\textwidth]{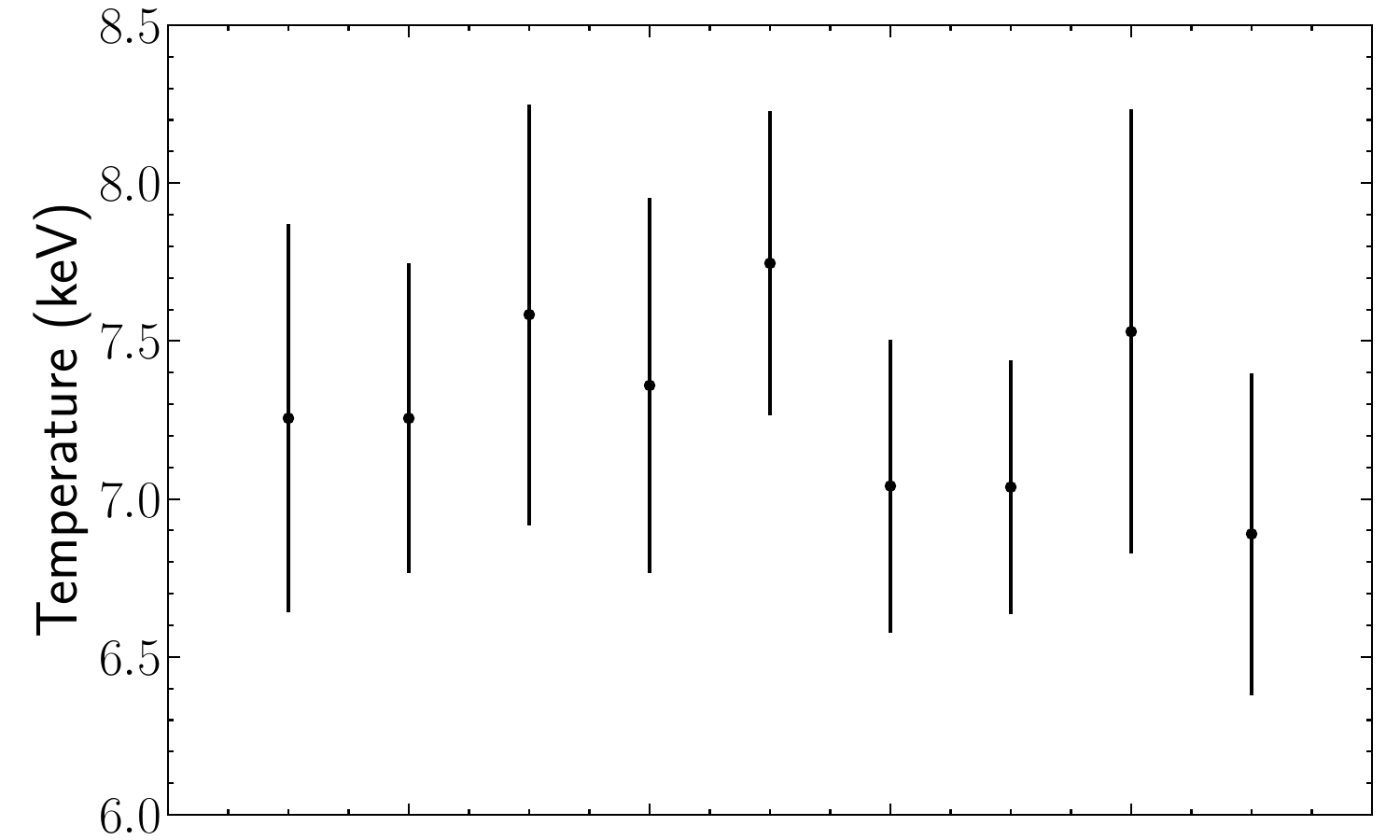}\\
\includegraphics[width=0.46\textwidth]{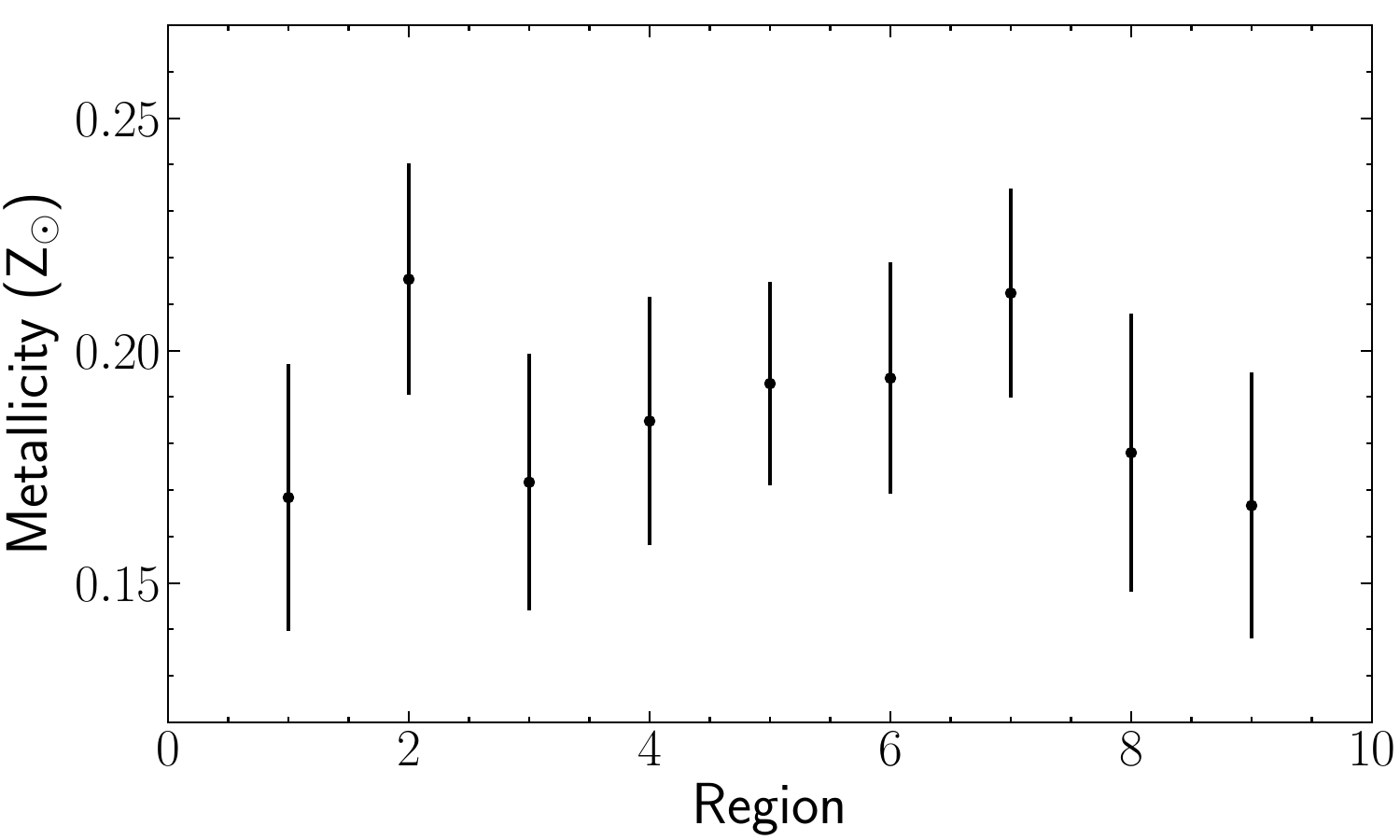} 
        \caption{Temperature (top panel) and metallicity (bottom panel) profiles obtained from the best fit results for manually selected regions following the A3266 merging structure. The spectral fitting is done in the $4-9.25$~keV energy range.} \label{fig_fit_sur2} 
\end{figure}

 We manually selected multiple regions based on the fractional difference in 0.5-2 keV X-ray surface brightness to study the velocity structure. 
 The top panel of Figure~\ref{fig_fit_sur1} shows the regions used for the analysis, which are similar to those identified in \citet{san22}. 
 The regions can be identified as follows: the central core (region 1), cavities (regions 2, 3, 4), the W structure (region 5), the NE structures (regions 6, 7), the bridge (region 8), and the NW structure (region 9). 
Table~\ref{tab_merging_fits} shows the best-fit results.
As shown in the spectral map (see subsection~\ref{spec_fit_sec}), we found redshifted velocities for all regions, with the largest velocities corresponding to cavity 2 (\(626 \pm 284\) km/s) and the NE structure (region 6, \(596 \pm 274\) km/s). 
The velocities for the remaining structures were similar, considering the uncertainties. 
We have also included the mean optical redshift for galaxies from \citet{deh17} within each one of the regions obtained.
The mean galactic velocities are within the ICM velocities except for regions 1, 4, 7, and 8.
The mean ICM velocity for all regions is $\sim 275$~km/s, and the mean galactic velocities are $\sim 525$~km/s.
However, the regions following the surface brightness discontinuities do not correspond to the sub-structures identified in \citet{deh17}.
Figure~\ref{fig_fit_sur2} shows the temperature (top panel) and the metallicity (bottom panel) for these regions. 
Both parameters tend to be uniform across the different regions, with a hint of increasing metallicity in region 2.

 \begin{table}
\scriptsize 
\caption{\label{tab_merging_fits}A3266 cluster best-fit parameters for manually selected regions following the A3266 merging structure. }
\centering
\begin{tabular}{ccccccc}
\\
Region &\multicolumn{5}{c}{{\tt apec} model}  \\
\hline
 &$kT$ & Z& $z$   & $norm$   & cstat/dof\\ 
  &   & &  ($\times 10^{-3}$) &   ($\times 10^{-3}$)  \\   
1&$7.26_{-0.51}^{+0.72}$&$0.17\pm 0.03$&$19.03_{-0.75}^{+0.68}$&$59.09_{-2.75}^{+2.81}$&$1455/1199$\\
2&$7.26_{-0.45}^{+0.53}$&$0.22\pm 0.03$&$18.87_{-0.56}^{+0.57}$&$61.62_{-1.77}^{+1.78}$&$1470/1199$\\
3&$7.58\pm 0.67$        &$0.17\pm 0.03$&$17.84_{-0.65}^{+0.68}$&$59.96_{-2.62}^{+2.72}$&$1478/1199$\\
4&$7.36_{-0.56}^{+0.63}$&$0.18\pm 0.03$&$15.62_{-0.53}^{+0.55}$&$59.97_{-2.07}^{+2.16}$&$1494/1199$\\
5&$7.75\pm 0.48$        &$0.19\pm 0.02$&$24.54_{-0.65}^{+0.68}$&$60.27_{-1.91}^{+1.93}$&$1489/1199$\\
6&$7.04_{-0.35}^{+0.57}$&$0.19\pm 0.03$&$15.69_{-0.51}^{+0.42}$&$61.51_{-2.07}^{+2.10}$&$1461/1199$\\
7&$7.04_{-0.32}^{+0.49}$&$0.21\pm 0.02$&$18.16_{-0.50}^{+0.43}$&$60.51_{-1.71}^{+1.66}$&$1536/1199$\\
8&$7.53\pm 0.70$        &$0.18\pm 0.03$&$17.78_{-0.68}^{+0.72}$&$60.39_{-2.72}^{+2.77}$&$1466/1199$\\
9&$6.89_{-0.25}^{+0.77}$&$0.17\pm 0.03$&$15.54_{-0.65}^{+0.38}$&$59.56_{-2.77}^{+2.86}$&$1473/1199$\\
\\ 
 \hline
\end{tabular}
\end{table}

\subsection{Probability distribution function}

\begin{figure} 
\centering  
\includegraphics[width=0.46\textwidth]{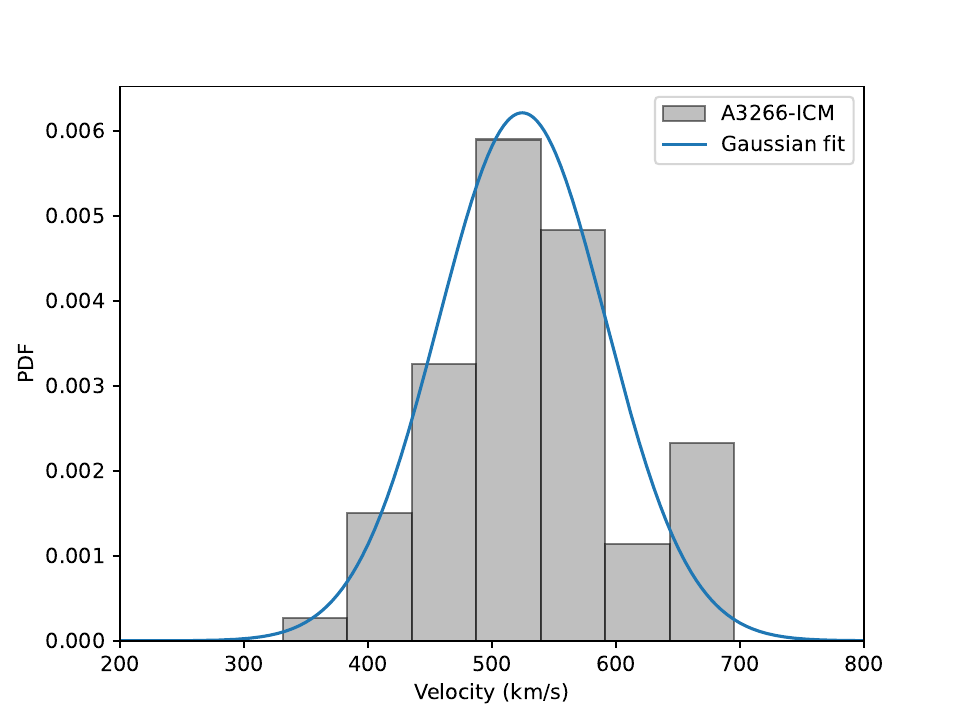} 
\caption{\emph{Top panel} Velocity PDF of the hot ICM obtained for the A3266 velocity map. The blue line corresponds to the best-fit using a Gaussian model.}\label{fig_pdf_vsf} 
\end{figure}

Figure~\ref{fig_pdf_vsf} shows the velocity probability distribution function (PDF) computed from the velocity map, weighted by area.
We computed normality tests, including the Shapiro-Wilk \citep[][SHA+65]{sha65} and D'Agostino and Pearson's \citep[][DAG+73]{dag73}, to determine if a Gaussian well models the data.
We found that the PDF follows a normal distribution as the $p$-value is larger than the $\alpha=0.05$ level for both tests (0.63 for SHA+65 and 0.71 for DAG+73).
The overall redshift obtained from the Gaussian fit is $524\pm 21$~km/s with velocity dispersion of $\sigma_{v}=68\pm 12$~km/s.
 
\subsection{Velocity structure function}\label{sec_vsf} 
\begin{figure}
\centering   
\includegraphics[width=0.49\textwidth]{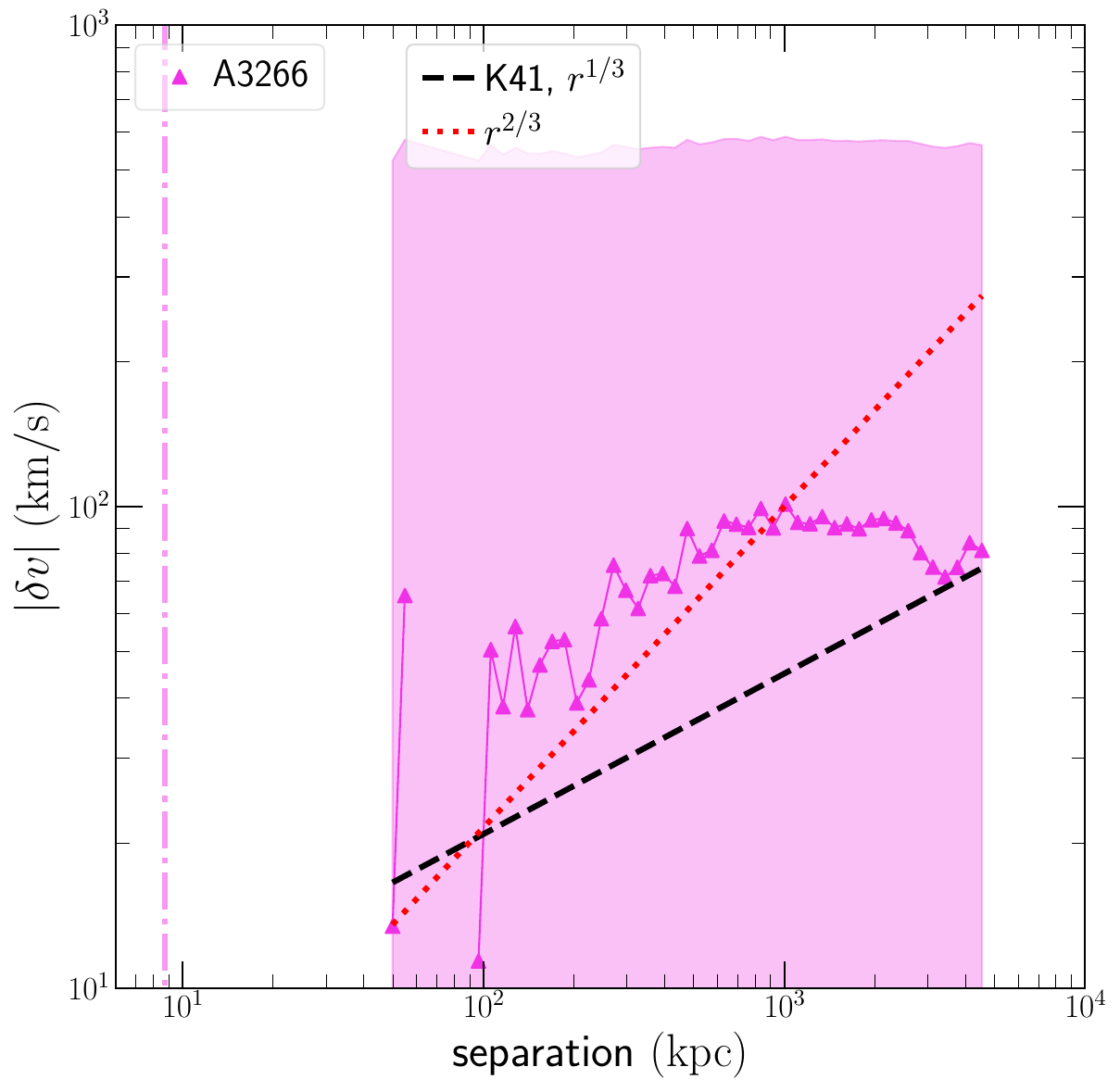} 
\caption{VSFs of the hot ICM within A3266. Power laws with slope $\sim$1/3 (i.e., Kolmogorov) and $\sim$2/3 are also included. The shaded region shows the $1-\sigma$ interval around the mean VSF. The vertical `dashdot' line indicates the PSF scale.}\label{fig_vsf_plots} 
\end{figure}

Following \citet{gat23c}, we computed the first-order  velocity structure function by taking the weighted average of the difference between the line-of-sight velocities ($v$) of two points separated by $r$. 
Mathematically, we define it as:
\begin{equation}
    \delta v (r) = \frac{\sum_{\mathbf{x}}{w(\mathbf{x}+\mathbf{e_1}r, \mathbf{x}) |v(\mathbf{x}+\mathbf{e_1}r)-v(\mathbf{x})|}}{\sum_\mathbf{x}{w(\mathbf{x}+\mathbf{e_1}r, \mathbf{x})}}
\end{equation}
where $x$ denotes the position of any point in the dataset and $\mathbf{e_1}$ denotes a unit-vector in any direction. We bin $\delta v$ into logarithmically-spaced bins of separation $r$.
Figure~\ref{fig_vsf_plots} shows the VSF weighted by the velocity uncertainties.
The uncertainties are large, as the velocity differences are smaller than the noise level.

\section{Discussion}\label{sec_dis}
We have found that the overall gas moves away from the observer throughout the entire system  (Figure~\ref{fig_spec_map1}), relative to the cluster redshift, with velocities increasing from $\sim 300-800$ km/s.
After subtracting the average redshift, the velocity spectral map  does show an opposite blueshifted-redshifted structure around the cluster center (Figure~\ref{fig_spec_rest}), with the blueshifted region displaying a lower metallicity.
However, there is no clear spiral pattern associated with gas sloshing \citep[see, for example, Figure~18 in][]{san20}. 
\citet{deh17} found that the A3266 core has a velocity dispersion of $\sim 1460$~km/s and that the dispersion of the whole cluster is $\sim 1340$~km/s in their analysis of spectroscopic redshifts of sources within the cluster. 
They concluded that there is a range of continuous dynamical interactions taking place within this system.

We did not find hints for a multimodal PDF, compared, for example, with the Ophiuchus cluster \citep{gat23c}.
Simulations predict a $\sigma_{v}$ value in the  $\sim 200-400$~km/s range with indications of isotropy for the hot ICM phase \citep{ayr23}.
For the Perseus cluster, {\it Hitomi} measured up to $\sigma_{v}\sim220$~km/s, while {\it XMM-Newton} measurements by \citep{gat23c} show a velocity dispersion of $\sigma_{v}=344\pm 18$~km/s (Virgo), $\sigma_{v}=371\pm 12$~km/s (Centaurus) and $\sigma_{v}=847\pm 92$~km/s (Ophiuchus).
Whereas the Perseus cluster velocity dispersion measured by {\it Hitomi} was due to line broadening, the {\it XMM-Newton} measurements provide dispersion velocities from the range of shifts.
The VSF results, on the other hand, can be interpreted as upper limits $<600$~km/s up to $\sim 3000$~kpc scales, in good agreement with the VSFs computed for the Virgo, Centaurus and Ophiuchus clusters \citep[see Figure 3 in ][]{gat23c}.

\citet{deh17} measured an overall optical redshift of the entire cluster and related structures to be $\sim 200\pm 50$~km/s with respect to the cluster core with velocity dispersion of $1337\pm 67$~km/s, such high dispersion does not allow a direct comparison with the hot ICM velocity distribution.
However, we note that they cover more considerable distances than our analysis.
They indicated that such high-velocity dispersion is not due to virialized motions but rather due to relative motions of the cluster components.
For example, the western core component substructure has a redshift of $\sim 600$ km/s with respect to the cluster core.
Other redshifted sub-structures in their analysis include the subclusters 2 and 3, with velocities of $\sim 559$~km/s and $\sim 393$~km/s, respectively \citep[][see their Table 3]{deh17}.
Large velocities can be observed due to subgroups, strong merger shocks, and/or substructures \citep{ang15,ang16,ota16,liu18}.

The most exciting feature observed in our analysis is the mean velocity of the hot gas, redshifted with respect to the A3266 cluster, measured at $\sim 300$~km/s (see Sections~\ref{spec_fit_sec},~\ref{fit_rings} and~\ref{sub_sur_fits}).
Moreover, the mean optical velocity values tend to be $>2$ times larger than the mean ICM velocities for the different regions analyzed (i.e., the spectral map and the manually selected extraction regions). 
As indicated by the optical redshifts measured by \citet{deh17}, the offset between the ICM and the cluster is due mainly to substructures in the cluster associated with the merging.

\section{Conclusions and summary}\label{sec_con}
We have analyzed the velocity structure of the hot ICM within the A3266 galaxy cluster.
We measured line-of-sight velocities of the gas using the technique developed by \citet{san20,gat22a,gat22b}, which allows a better calibration of the absolute energy scale of the {\it XMM-Newton} EPIC-pn detector.
Our main findings and conclusions are:  
 \begin{enumerate}
\item We studied the morphological structures within the galaxy cluster by applying a GGM and adaptatively smoothed GGM filter to the EPIC-pn X-ray image.
\item We made 2D projected maps for the velocity, temperature, and metallicity. We have found that the hot gas within the system displays a redshifted velocity with respect to the cluster redshift along all the fields of view. After subtracting the average redshift, the velocity spectral map does show an opposite blueshifted-redshifted structure, with the blueshifted region displaying a lower metallicity.
\item We studied the velocity distribution by creating non-overlapping circular regions. Our results are consistent with redshifted velocities as far as 1125 kpc from the cluster core.
\item We analyzed the velocity distribution along regions following the discontinuities in the surface brightness. We found redshifted velocities for all regions with the largest velocities to be $805\pm 284$~km/s.
\item We compute the velocity PDF from the velocity map. By applying a normality test, we found that the PDF follows an unimodal normal distribution, as predicted by simulations.
\item We obtained upper limits for the VSF of $< 600$~km/s up to $\sim 3000$~kpc scales. 
\end{enumerate}  

Finally, the present work will be followed by a detailed study of the temperature and abundance distribution by modeling the soft X-ray spectra of the A3266 galaxy cluster.

\begin{acknowledgements} 
This work was supported by the Deutsche Zentrum f\"ur Luft- und Raumfahrt (DLR) under the Verbundforschung programme (Messung von Schwapp-, Verschmelzungs- und R\"uckkopplungsgeschwindigkeiten in Galaxienhaufen.). 
This work is based on observations obtained with XMM-Newton, an ESA science mission with instruments and contributions directly funded by ESA Member States and NASA. 
This research was carried out on the High Performance Computing resources of the raven cluster at the Max Planck Computing and Data Facility (MPCDF) in Garching operated by the Max Planck Society (MPG).  
\end{acknowledgements}

\bibliographystyle{aa}

\begin{thebibliography}{68}
\expandafter\ifx\csname natexlab\endcsname\relax\def\natexlab#1{#1}\fi

\bibitem[{{Ahoranta} {et~al.}(2016){Ahoranta}, {Finoguenov}, {Pinto},
  {Sanders}, {Kaastra}, {de Plaa}, \& {Fabian}}]{aho16}
{Ahoranta}, J., {Finoguenov}, A., {Pinto}, C., {et~al.} 2016, \aap, 592, A145

\bibitem[{{Ascasibar} \& {Markevitch}(2006)}]{asc06}
{Ascasibar}, Y. \& {Markevitch}, M. 2006, \apj, 650, 102

\bibitem[{{Ayromlou} {et~al.}(2023){Ayromlou}, {Nelson}, {Pillepich}, {Rohr},
  {Truong}, {Li}, {Simionescu}, {Lehle}, \& {Lee}}]{ayr23}
{Ayromlou}, M., {Nelson}, D., {Pillepich}, A., {et~al.} 2023, arXiv e-prints,
  arXiv:2311.06339

\bibitem[{{Cash}(1979)}]{cas79}
{Cash}, W. 1979, \apj, 228, 939

\bibitem[{D'Agostino \& Pearson(1973)}]{dag73}
D'Agostino, R. \& Pearson, E.~S. 1973, 60, 613, see correction
  \cite{DAgostino:1974:CAT}.

\bibitem[{{Dehghan} {et~al.}(2017){Dehghan}, {Johnston-Hollitt}, {Colless}, \&
  {Miller}}]{deh17}
{Dehghan}, S., {Johnston-Hollitt}, M., {Colless}, M., \& {Miller}, R. 2017,
  \mnras, 468, 2645

\bibitem[{{Dennerl} {et~al.}(2020){Dennerl}, {Andritschke}, {Br{\"a}uninger},
  {Burkert}, {Burwitz}, {Emberger}, {Freyberg}, {Friedrich}, {Gaida},
  {Granato}, {Hartner}, {von Kienlin}, {Meidinger}, {Menz}, \&
  {Predehl}}]{den20}
{Dennerl}, K., {Andritschke}, R., {Br{\"a}uninger}, H., {et~al.} 2020, in
  Society of Photo-Optical Instrumentation Engineers (SPIE) Conference Series,
  Vol. 11444, Space Telescopes and Instrumentation 2020: Ultraviolet to Gamma
  Ray, ed. J.-W.~A. {den Herder}, S.~{Nikzad}, \& K.~{Nakazawa}, 114444Q

\bibitem[{{Edge} {et~al.}(1990){Edge}, {Stewart}, {Fabian}, \&
  {Arnaud}}]{edg90}
{Edge}, A.~C., {Stewart}, G.~C., {Fabian}, A.~C., \& {Arnaud}, K.~A. 1990,
  \mnras, 245, 559

\bibitem[{{Ettori} {et~al.}(2019){Ettori}, {Ghirardini}, {Eckert},
  {Pointecouteau}, {Gastaldello}, {Sereno}, {Gaspari}, {Ghizzardi},
  {Roncarelli}, \& {Rossetti}}]{ett19}
{Ettori}, S., {Ghirardini}, V., {Eckert}, D., {et~al.} 2019, \aap, 621, A39

\bibitem[{{Fabian} {et~al.}(2017){Fabian}, {Walker}, {Russell}, {Pinto},
  {Sanders}, \& {Reynolds}}]{fab17}
{Fabian}, A.~C., {Walker}, S.~A., {Russell}, H.~R., {et~al.} 2017, \mnras, 464,
  L1

\bibitem[{{Federrath} {et~al.}(2021){Federrath}, {Klessen}, {Iapichino}, \&
  {Beattie}}]{fed21}
{Federrath}, C., {Klessen}, R.~S., {Iapichino}, L., \& {Beattie}, J.~R. 2021,
  Nature Astronomy, 5, 365

\bibitem[{{Federrath} {et~al.}(2010){Federrath}, {Roman-Duval}, {Klessen},
  {Schmidt}, \& {Mac Low}}]{fed10}
{Federrath}, C., {Roman-Duval}, J., {Klessen}, R.~S., {Schmidt}, W., \& {Mac
  Low}, M.~M. 2010, \aap, 512, A81

\bibitem[{{Finoguenov} {et~al.}(2006){Finoguenov}, {Henriksen}, {Miniati},
  {Briel}, \& {Jones}}]{fin06}
{Finoguenov}, A., {Henriksen}, M.~J., {Miniati}, F., {Briel}, U.~G., \&
  {Jones}, C. 2006, \apj, 643, 790

\bibitem[{{Foster} {et~al.}(2019){Foster}, {Smith}, \& {Brickhouse}}]{fos19}
{Foster}, A., {Smith}, R., \& {Brickhouse}, N.~S. 2019, in American
  Astronomical Society Meeting Abstracts, Vol. 233, American Astronomical
  Society Meeting Abstracts \#233, 251.05

\bibitem[{{Gaspari} {et~al.}(2014){Gaspari}, {Churazov}, {Nagai}, {Lau}, \&
  {Zhuravleva}}]{gas14}
{Gaspari}, M., {Churazov}, E., {Nagai}, D., {Lau}, E.~T., \& {Zhuravleva}, I.
  2014, \aap, 569, A67

\bibitem[{{Gatuzz} {et~al.}(2023{\natexlab{a}}){Gatuzz}, {Mohapatra},
  {Federrath}, {Sanders}, {Liu}, {Walker}, \& {Pinto}}]{gat23c}
{Gatuzz}, E., {Mohapatra}, R., {Federrath}, C., {et~al.} 2023{\natexlab{a}},
  \mnras, 524, 2945

\bibitem[{{Gatuzz} {et~al.}(2022{\natexlab{a}}){Gatuzz}, {Sanders}, {Canning},
  {Dennerl}, {Fabian}, {Pinto}, {Russell}, {Tamura}, {Walker}, \&
  {ZuHone}}]{gat22b}
{Gatuzz}, E., {Sanders}, J.~S., {Canning}, R., {et~al.} 2022{\natexlab{a}},
  \mnras, 513, 1932

\bibitem[{{Gatuzz} {et~al.}(2023{\natexlab{b}}){Gatuzz}, {Sanders}, {Dennerl},
  {Liu}, {Fabian}, {Pinto}, {Eckert}, {Russell}, {Tamura}, {Walker}, \&
  {ZuHone}}]{gat23b}
{Gatuzz}, E., {Sanders}, J.~S., {Dennerl}, K., {et~al.} 2023{\natexlab{b}},
  \mnras, 520, 4793

\bibitem[{{Gatuzz} {et~al.}(2023{\natexlab{c}}){Gatuzz}, {Sanders}, {Dennerl},
  {Liu}, {Fabian}, {Pinto}, {Eckert}, {Russell}, {Tamura}, {Walker}, \&
  {ZuHone}}]{gat23a}
{Gatuzz}, E., {Sanders}, J.~S., {Dennerl}, K., {et~al.} 2023{\natexlab{c}},
  arXiv e-prints, arXiv:2303.17556

\bibitem[{{Gatuzz} {et~al.}(2023{\natexlab{d}}){Gatuzz}, {Sanders}, {Dennerl},
  {Liu}, {Fabian}, {Pinto}, {Eckert}, {Walker}, \& {ZuHone}}]{gat23d}
{Gatuzz}, E., {Sanders}, J.~S., {Dennerl}, K., {et~al.} 2023{\natexlab{d}},
  \mnras, 525, 6394

\bibitem[{{Gatuzz} {et~al.}(2023{\natexlab{e}}){Gatuzz}, {Sanders}, {Dennerl},
  {Liu}, {Fabian}, {Pinto}, {Eckert}, {Walker}, \& {ZuHone}}]{gat23e}
{Gatuzz}, E., {Sanders}, J.~S., {Dennerl}, K., {et~al.} 2023{\natexlab{e}},
  \mnras, 526, 396

\bibitem[{{Gatuzz} {et~al.}(2022{\natexlab{b}}){Gatuzz}, {Sanders}, {Dennerl},
  {Pinto}, {Fabian}, {Tamura}, {Walker}, \& {ZuHone}}]{gat22a}
{Gatuzz}, E., {Sanders}, J.~S., {Dennerl}, K., {et~al.} 2022{\natexlab{b}},
  \mnras, 511, 4511

\bibitem[{{Genel} {et~al.}(2014){Genel}, {Vogelsberger}, {Springel}, {Sijacki},
  {Nelson}, {Snyder}, {Rodriguez-Gomez}, {Torrey}, \& {Hernquist}}]{gen14}
{Genel}, S., {Vogelsberger}, M., {Springel}, V., {et~al.} 2014, \mnras, 445,
  175

\bibitem[{{Ghirardini} {et~al.}(2019){Ghirardini}, {Eckert}, {Ettori},
  {Pointecouteau}, {Molendi}, {Gaspari}, {Rossetti}, {De Grandi}, {Roncarelli},
  {Bourdin}, {Mazzotta}, {Rasia}, \& {Vazza}}]{ghi19}
{Ghirardini}, V., {Eckert}, D., {Ettori}, S., {et~al.} 2019, \aap, 621, A41

\bibitem[{{Henriksen} {et~al.}(2000){Henriksen}, {Donnelly}, \&
  {Davis}}]{hen00}
{Henriksen}, M., {Donnelly}, R.~H., \& {Davis}, D.~S. 2000, \apj, 529, 692

\bibitem[{{Henriksen} \& {Tittley}(2002)}]{hen02}
{Henriksen}, M.~J. \& {Tittley}, E.~R. 2002, \apj, 577, 701

\bibitem[{{Hitomi Collaboration} {et~al.}(2016){Hitomi Collaboration},
  {Aharonian}, {Akamatsu}, {Akimoto}, {Allen}, {Anabuki}, {Angelini}, {Arnaud},
  {Audard}, {Awaki}, {Axelsson}, {Bamba}, {Bautz}, {Blandford}, {Brenneman},
  {Brown}, {Bulbul}, {Cackett}, {Chernyakova}, {Chiao}, {Coppi}, {Costantini},
  {de Plaa}, {den Herder}, {Done}, {Dotani}, {Ebisawa}, {Eckart}, {Enoto},
  {Ezoe}, {Fabian}, {Ferrigno}, {Foster}, {Fujimoto}, {Fukazawa}, {Furuzawa},
  {Galeazzi}, {Gallo}, {Gandhi}, {Giustini}, {Goldwurm}, {Gu}, {Guainazzi},
  {Haba}, {Hagino}, {Hamaguchi}, {Harrus}, {Hatsukade}, {Hayashi}, {Hayashi},
  {Hayashida}, {Hiraga}, {Hornschemeier}, {Hoshino}, {Hughes}, {Iizuka},
  {Inoue}, {Inoue}, {Ishibashi}, {Ishida}, {Ishikawa}, {Ishisaki}, {Itoh},
  {Iyomoto}, {Kaastra}, {Kallman}, {Kamae}, {Kara}, {Kataoka}, {Katsuda},
  {Katsuta}, {Kawaharada}, {Kawai}, {Kelley}, {Khangulyan}, {Kilbourne},
  {King}, {Kitaguchi}, {Kitamoto}, {Kitayama}, {Kohmura}, {Kokubun}, {Koyama},
  {Koyama}, {Kretschmar}, {Krimm}, {Kubota}, {Kunieda}, {Laurent}, {Lebrun},
  {Lee}, {Leutenegger}, {Limousin}, {Loewenstein}, {Long}, {Lumb}, {Madejski},
  {Maeda}, {Maier}, {Makishima}, {Markevitch}, {Matsumoto}, {Matsushita},
  {McCammon}, {McNamara}, {Mehdipour}, {Miller}, {Miller}, {Mineshige},
  {Mitsuda}, {Mitsuishi}, {Miyazawa}, {Mizuno}, {Mori}, {Mori}, {Moseley},
  {Mukai}, {Murakami}, {Murakami}, {Mushotzky}, {Nagino}, {Nakagawa},
  {Nakajima}, {Nakamori}, {Nakano}, {Nakashima}, {Nakazawa}, {Nobukawa},
  {Noda}, {Nomachi}, {O'Dell}, {Odaka}, {Ohashi}, {Ohno}, {Okajima}, {Ota},
  {Ozaki}, {Paerels}, {Paltani}, {Parmar}, {Petre}, {Pinto}, {Pohl}, {Porter},
  {Pottschmidt}, {Ramsey}, {Reynolds}, {Russell}, {Safi-Harb}, {Saito},
  {Sakai}, {Sameshima}, {Sato}, {Sato}, {Sato}, {Sawada}, {Schartel},
  {Serlemitsos}, {Seta}, {Shidatsu}, {Simionescu}, {Smith}, {Soong}, {Stawarz},
  {Sugawara}, {Sugita}, {Szymkowiak}, {Tajima}, {Takahashi}, {Takahashi},
  {Takeda}, {Takei}, {Tamagawa}, {Tamura}, {Tamura}, {Tanaka}, {Tanaka},
  {Tanaka}, {Tashiro}, {Tawara}, {Terada}, {Terashima}, {Tombesi}, {Tomida},
  {Tsuboi}, {Tsujimoto}, {Tsunemi}, {Tsuru}, {Uchida}, {Uchiyama}, {Uchiyama},
  {Ueda}, {Ueda}, {Ueno}, {Uno}, {Urry}, {Ursino}, {de Vries}, {Watanabe},
  {Werner}, {Wik}, {Wilkins}, {Williams}, {Yamada}, {Yamaguchi}, {Yamaoka},
  {Yamasaki}, {Yamauchi}, {Yamauchi}, {Yaqoob}, {Yatsu}, {Yonetoku}, {Yoshida},
  {Yuasa}, {Zhuravleva}, \& {Zoghbi}}]{hit16}
{Hitomi Collaboration}, {Aharonian}, F., {Akamatsu}, H., {et~al.} 2016, \nat,
  535, 117

\bibitem[{{Hitomi Collaboration} {et~al.}(2018){Hitomi Collaboration},
  {Aharonian}, {Akamatsu}, {Akimoto}, {Allen}, {Angelini}, {Audard}, {Awaki},
  {Axelsson}, {Bamba}, {Bautz}, {Blandford}, {Brenneman}, {Brown}, {Bulbul},
  {Cackett}, {Chernyakova}, {Chiao}, {Coppi}, {Costantini}, {de Plaa}, {de
  Vries}, {den Herder}, {Done}, {Dotani}, {Ebisawa}, {Eckart}, {Enoto}, {Ezoe},
  {Fabian}, {Ferrigno}, {Foster}, {Fujimoto}, {Fukazawa}, {Furuzawa},
  {Galeazzi}, {Gallo}, {Gandhi}, {Giustini}, {Goldwurm}, {Gu}, {Guainazzi},
  {Haba}, {Hagino}, {Hamaguchi}, {Harrus}, {Hatsukade}, {Hayashi}, {Hayashi},
  {Hayashida}, {Hell}, {Hiraga}, {Hornschemeier}, {Hoshino}, {Hughes},
  {Ichinohe}, {Iizuka}, {Inoue}, {Inoue}, {Ishida}, {Ishikawa}, {Ishisaki},
  {Iwai}, {Kaastra}, {Kallman}, {Kamae}, {Kataoka}, {Katsuda}, {Kawai},
  {Kelley}, {Kilbourne}, {Kitaguchi}, {Kitamoto}, {Kitayama}, {Kohmura},
  {Kokubun}, {Koyama}, {Koyama}, {Kretschmar}, {Krimm}, {Kubota}, {Kunieda},
  {Laurent}, {Lee}, {Leutenegger}, {Limousin}, {Loewenstein}, {Long}, {Lumb},
  {Madejski}, {Maeda}, {Maier}, {Makishima}, {Markevitch}, {Matsumoto},
  {Matsushita}, {McCammon}, {McNamara}, {Mehdipour}, {Miller}, {Miller},
  {Mineshige}, {Mitsuda}, {Mitsuishi}, {Miyazawa}, {Mizuno}, {Mori}, {Mori},
  {Mukai}, {Murakami}, {Mushotzky}, {Nakagawa}, {Nakajima}, {Nakamori},
  {Nakashima}, {Nakazawa}, {Nobukawa}, {Nobukawa}, {Noda}, {Odaka}, {Ohashi},
  {Ohno}, {Okajima}, {Ota}, {Ozaki}, {Paerels}, {Paltani}, {Petre}, {Pinto},
  {Porter}, {Pottschmidt}, {Reynolds}, {Safi-Harb}, {Saito}, {Sakai}, {Sasaki},
  {Sato}, {Sato}, {Sato}, {Sawada}, {Schartel}, {Serlemtsos}, {Seta},
  {Shidatsu}, {Simionescu}, {Smith}, {Soong}, {Stawarz}, {Sugawara}, {Sugita},
  {Szymkowiak}, {Tajima}, {Takahashi}, {Takahashi}, {Takeda}, {Takei},
  {Tamagawa}, {Tamura}, {Tanaka}, {Tanaka}, {Tanaka}, {Tashiro}, {Tawara},
  {Terada}, {Terashima}, {Tombesi}, {Tomida}, {Tsuboi}, {Tsujimoto}, {Tsunemi},
  {Tsuru}, {Uchida}, {Uchiyama}, {Uchiyama}, {Ueda}, {Ueda}, {Uno}, {Urry},
  {Ursino}, {Watanabe}, {Werner}, {Wilkins}, {Williams}, {Yamada}, {Yamaguchi},
  {Yamaoka}, {Yamasaki}, {Yamauchi}, {Yamauchi}, {Yaqoob}, {Yatsu}, {Yonetoku},
  {Zhuravleva}, {Zoghbi}, \& {Raassen}}]{hit18}
{Hitomi Collaboration}, {Aharonian}, F., {Akamatsu}, H., {et~al.} 2018, \pasj,
  70, 12

\bibitem[{{Hofmann} {et~al.}(2016){Hofmann}, {Sanders}, {Nandra}, {Clerc}, \&
  {Gaspari}}]{hof16}
{Hofmann}, F., {Sanders}, J.~S., {Nandra}, K., {Clerc}, N., \& {Gaspari}, M.
  2016, \aap, 585, A130

\bibitem[{{Kalberla} {et~al.}(2005){Kalberla}, {Burton}, {Hartmann}, {Arnal},
  {Bajaja}, {Morras}, \& {P{\"o}ppel}}]{kal05}
{Kalberla}, P.~M.~W., {Burton}, W.~B., {Hartmann}, D., {et~al.} 2005, \aap,
  440, 775

\bibitem[{{Lau} {et~al.}(2009){Lau}, {Kravtsov}, \& {Nagai}}]{lau09}
{Lau}, E.~T., {Kravtsov}, A.~V., \& {Nagai}, D. 2009, \apj, 705, 1129

\bibitem[{{Liu} {et~al.}(2018){Liu}, {Yu}, {Diaferio}, {Tozzi}, {Hwang},
  {Umetsu}, {Okabe}, \& {Yang}}]{liu18}
{Liu}, A., {Yu}, H., {Diaferio}, A., {et~al.} 2018, \apj, 863, 102

\bibitem[{{Liu} {et~al.}(2015){Liu}, {Yu}, {Tozzi}, \& {Zhu}}]{ang15}
{Liu}, A., {Yu}, H., {Tozzi}, P., \& {Zhu}, Z.-H. 2015, \apj, 809, 27

\bibitem[{{Liu} {et~al.}(2016){Liu}, {Yu}, {Tozzi}, \& {Zhu}}]{ang16}
{Liu}, A., {Yu}, H., {Tozzi}, P., \& {Zhu}, Z.-H. 2016, \apj, 821, 29

\bibitem[{{Lodders} {et~al.}(2009){Lodders}, {Palme}, \& {Gail}}]{lod09}
{Lodders}, K., {Palme}, H., \& {Gail}, H.-P. 2009, Landolt B{\"o}rnstein
  [\eprint[arXiv]{0901.1149}]

\bibitem[{{Marinacci} {et~al.}(2018){Marinacci}, {Vogelsberger}, {Pakmor},
  {Torrey}, {Springel}, {Hernquist}, {Nelson}, {Weinberger}, {Pillepich},
  {Naiman}, \& {Genel}}]{mar18a}
{Marinacci}, F., {Vogelsberger}, M., {Pakmor}, R., {et~al.} 2018, \mnras, 480,
  5113

\bibitem[{{Naiman} {et~al.}(2018){Naiman}, {Pillepich}, {Springel},
  {Ramirez-Ruiz}, {Torrey}, {Vogelsberger}, {Pakmor}, {Nelson}, {Marinacci},
  {Hernquist}, {Weinberger}, \& {Genel}}]{nai18}
{Naiman}, J.~P., {Pillepich}, A., {Springel}, V., {et~al.} 2018, \mnras, 477,
  1206

\bibitem[{{Navarro} {et~al.}(1996){Navarro}, {Frenk}, \& {White}}]{nav96}
{Navarro}, J.~F., {Frenk}, C.~S., \& {White}, S. D.~M. 1996, \apj, 462, 563

\bibitem[{{Nelson} {et~al.}(2015){Nelson}, {Pillepich}, {Genel},
  {Vogelsberger}, {Springel}, {Torrey}, {Rodriguez-Gomez}, {Sijacki}, {Snyder},
  {Griffen}, {Marinacci}, {Blecha}, {Sales}, {Xu}, \& {Hernquist}}]{nel15}
{Nelson}, D., {Pillepich}, A., {Genel}, S., {et~al.} 2015, Astronomy and
  Computing, 13, 12

\bibitem[{{Nelson} {et~al.}(2018){Nelson}, {Pillepich}, {Springel},
  {Weinberger}, {Hernquist}, {Pakmor}, {Genel}, {Torrey}, {Vogelsberger},
  {Kauffmann}, {Marinacci}, \& {Naiman}}]{nel18}
{Nelson}, D., {Pillepich}, A., {Springel}, V., {et~al.} 2018, \mnras, 475, 624

\bibitem[{{Ota} \& {Yoshida}(2016)}]{ota16}
{Ota}, N. \& {Yoshida}, H. 2016, \pasj, 68, S19

\bibitem[{{Pillepich} {et~al.}(2018{\natexlab{a}}){Pillepich}, {Nelson},
  {Hernquist}, {Springel}, {Pakmor}, {Torrey}, {Weinberger}, {Genel}, {Naiman},
  {Marinacci}, \& {Vogelsberger}}]{mar18b}
{Pillepich}, A., {Nelson}, D., {Hernquist}, L., {et~al.} 2018{\natexlab{a}},
  \mnras, 475, 648

\bibitem[{{Pillepich} {et~al.}(2018{\natexlab{b}}){Pillepich}, {Springel},
  {Nelson}, {Genel}, {Naiman}, {Pakmor}, {Hernquist}, {Torrey}, {Vogelsberger},
  {Weinberger}, \& {Marinacci}}]{pil18}
{Pillepich}, A., {Springel}, V., {Nelson}, D., {et~al.} 2018{\natexlab{b}},
  \mnras, 473, 4077

\bibitem[{{Pinto} {et~al.}(2015){Pinto}, {Sanders}, {Werner}, {de Plaa},
  {Fabian}, {Zhang}, {Kaastra}, {Finoguenov}, \& {Ahoranta}}]{pin15}
{Pinto}, C., {Sanders}, J.~S., {Werner}, N., {et~al.} 2015, \aap, 575, A38

\bibitem[{{Planck Collaboration} {et~al.}(2016){Planck Collaboration}, {Ade},
  {Aghanim}, {Arnaud}, {Ashdown}, {Aumont}, {Baccigalupi}, {Banday},
  {Barreiro}, {Bartlett}, {Bartolo}, {Battaner}, {Battye}, {Benabed},
  {Beno{\^\i}t}, {Benoit-L{\'e}vy}, {Bernard}, {Bersanelli}, {Bielewicz},
  {Bock}, {Bonaldi}, {Bonavera}, {Bond}, {Borrill}, {Bouchet}, {Boulanger},
  {Bucher}, {Burigana}, {Butler}, {Calabrese}, {Cardoso}, {Catalano},
  {Challinor}, {Chamballu}, {Chary}, {Chiang}, {Chluba}, {Christensen},
  {Church}, {Clements}, {Colombi}, {Colombo}, {Combet}, {Coulais}, {Crill},
  {Curto}, {Cuttaia}, {Danese}, {Davies}, {Davis}, {de Bernardis}, {de Rosa},
  {de Zotti}, {Delabrouille}, {D{\'e}sert}, {Di Valentino}, {Dickinson},
  {Diego}, {Dolag}, {Dole}, {Donzelli}, {Dor{\'e}}, {Douspis}, {Ducout},
  {Dunkley}, {Dupac}, {Efstathiou}, {Elsner}, {En{\ss}lin}, {Eriksen},
  {Farhang}, {Fergusson}, {Finelli}, {Forni}, {Frailis}, {Fraisse},
  {Franceschi}, {Frejsel}, {Galeotta}, {Galli}, {Ganga}, {Gauthier}, {Gerbino},
  {Ghosh}, {Giard}, {Giraud-H{\'e}raud}, {Giusarma}, {Gjerl{\o}w},
  {Gonz{\'a}lez-Nuevo}, {G{\'o}rski}, {Gratton}, {Gregorio}, {Gruppuso},
  {Gudmundsson}, {Hamann}, {Hansen}, {Hanson}, {Harrison}, {Helou},
  {Henrot-Versill{\'e}}, {Hern{\'a}ndez-Monteagudo}, {Herranz}, {Hildebrandt},
  {Hivon}, {Hobson}, {Holmes}, {Hornstrup}, {Hovest}, {Huang}, {Huffenberger},
  {Hurier}, {Jaffe}, {Jaffe}, {Jones}, {Juvela}, {Keih{\"a}nen}, {Keskitalo},
  {Kisner}, {Kneissl}, {Knoche}, {Knox}, {Kunz}, {Kurki-Suonio}, {Lagache},
  {L{\"a}hteenm{\"a}ki}, {Lamarre}, {Lasenby}, {Lattanzi}, {Lawrence}, {Leahy},
  {Leonardi}, {Lesgourgues}, {Levrier}, {Lewis}, {Liguori}, {Lilje},
  {Linden-V{\o}rnle}, {L{\'o}pez-Caniego}, {Lubin}, {Mac{\'\i}as-P{\'e}rez},
  {Maggio}, {Maino}, {Mandolesi}, {Mangilli}, {Marchini}, {Maris}, {Martin},
  {Martinelli}, {Mart{\'\i}nez-Gonz{\'a}lez}, {Masi}, {Matarrese}, {McGehee},
  {Meinhold}, {Melchiorri}, {Melin}, {Mendes}, {Mennella}, {Migliaccio},
  {Millea}, {Mitra}, {Miville-Desch{\^e}nes}, {Moneti}, {Montier}, {Morgante},
  {Mortlock}, {Moss}, {Munshi}, {Murphy}, {Naselsky}, {Nati}, {Natoli},
  {Netterfield}, {N{\o}rgaard-Nielsen}, {Noviello}, {Novikov}, {Novikov},
  {Oxborrow}, {Paci}, {Pagano}, {Pajot}, {Paladini}, {Paoletti}, {Partridge},
  {Pasian}, {Patanchon}, {Pearson}, {Perdereau}, {Perotto}, {Perrotta},
  {Pettorino}, {Piacentini}, {Piat}, {Pierpaoli}, {Pietrobon}, {Plaszczynski},
  {Pointecouteau}, {Polenta}, {Popa}, {Pratt}, {Pr{\'e}zeau}, {Prunet},
  {Puget}, {Rachen}, {Reach}, {Rebolo}, {Reinecke}, {Remazeilles}, {Renault},
  {Renzi}, {Ristorcelli}, {Rocha}, {Rosset}, {Rossetti}, {Roudier},
  {Rouill{\'e} d'Orfeuil}, {Rowan-Robinson}, {Rubi{\~n}o-Mart{\'\i}n},
  {Rusholme}, {Said}, {Salvatelli}, {Salvati}, {Sandri}, {Santos},
  {Savelainen}, {Savini}, {Scott}, {Seiffert}, {Serra}, {Shellard}, {Spencer},
  {Spinelli}, {Stolyarov}, {Stompor}, {Sudiwala}, {Sunyaev}, {Sutton},
  {Suur-Uski}, {Sygnet}, {Tauber}, {Terenzi}, {Toffolatti}, {Tomasi},
  {Tristram}, {Trombetti}, {Tucci}, {Tuovinen}, {T{\"u}rler}, {Umana},
  {Valenziano}, {Valiviita}, {Van Tent}, {Vielva}, {Villa}, {Wade}, {Wandelt},
  {Wehus}, {White}, {White}, {Wilkinson}, {Yvon}, {Zacchei}, \&
  {Zonca}}]{pla16}
{Planck Collaboration}, {Ade}, P.~A.~R., {Aghanim}, N., {et~al.} 2016, \aap,
  594, A13

\bibitem[{{Predehl} {et~al.}(2021){Predehl}, {Andritschke}, {Arefiev},
  {Babyshkin}, {Batanov}, {Becker}, {B{\"o}hringer}, {Bogomolov}, {Boller},
  {Borm}, {Bornemann}, {Br{\"a}uninger}, {Br{\"u}ggen}, {Brunner}, {Brusa},
  {Bulbul}, {Buntov}, {Burwitz}, {Burkert}, {Clerc}, {Churazov}, {Coutinho},
  {Dauser}, {Dennerl}, {Doroshenko}, {Eder}, {Emberger}, {Eraerds},
  {Finoguenov}, {Freyberg}, {Friedrich}, {Friedrich}, {F{\"u}rmetz},
  {Georgakakis}, {Gilfanov}, {Granato}, {Grossberger}, {Gueguen}, {Gureev},
  {Haberl}, {H{\"a}lker}, {Hartner}, {Hasinger}, {Huber}, {Ji}, {Kienlin},
  {Kink}, {Korotkov}, {Kreykenbohm}, {Lamer}, {Lomakin}, {Lapshov}, {Liu},
  {Maitra}, {Meidinger}, {Menz}, {Merloni}, {Mernik}, {Mican}, {Mohr},
  {M{\"u}ller}, {Nandra}, {Nazarov}, {Pacaud}, {Pavlinsky}, {Perinati},
  {Pfeffermann}, {Pietschner}, {Ramos-Ceja}, {Rau}, {Reiffers}, {Reiprich},
  {Robrade}, {Salvato}, {Sanders}, {Santangelo}, {Sasaki}, {Scheuerle},
  {Schmid}, {Schmitt}, {Schwope}, {Shirshakov}, {Steinmetz}, {Stewart},
  {Str{\"u}der}, {Sunyaev}, {Tenzer}, {Tiedemann}, {Tr{\"u}mper}, {Voron},
  {Weber}, {Wilms}, \& {Yaroshenko}}]{pre21}
{Predehl}, P., {Andritschke}, R., {Arefiev}, V., {et~al.} 2021, \aap, 647, A1

\bibitem[{{Sanders} {et~al.}(2022{\natexlab{a}}){Sanders}, {Biffi},
  {Br{\"u}ggen}, {Bulbul}, {Dennerl}, {Dolag}, {Erben}, {Freyberg}, {Gatuzz},
  {Ghirardini}, {Hoang}, {Klein}, {Liu}, {Merloni}, {Pacaud}, {Ramos-Ceja},
  {Reiprich}, \& {ZuHone}}]{san22}
{Sanders}, J.~S., {Biffi}, V., {Br{\"u}ggen}, M., {et~al.} 2022{\natexlab{a}},
  \aap, 661, A36

\bibitem[{{Sanders} {et~al.}(2022{\natexlab{b}}){Sanders}, {Biffi},
  {Br{\"u}ggen}, {Bulbul}, {Dennerl}, {Dolag}, {Erben}, {Freyberg}, {Gatuzz},
  {Ghirardini}, {Hoang}, {Klein}, {Liu}, {Merloni}, {Pacaud}, {Ramos-Ceja},
  {Reiprich}, \& {ZuHone}}]{san22a}
{Sanders}, J.~S., {Biffi}, V., {Br{\"u}ggen}, M., {et~al.} 2022{\natexlab{b}},
  \aap, 661, A36

\bibitem[{{Sanders} {et~al.}(2020){Sanders}, {Dennerl}, {Russell}, {Eckert},
  {Pinto}, {Fabian}, {Walker}, {Tamura}, {ZuHone}, \& {Hofmann}}]{san20}
{Sanders}, J.~S., {Dennerl}, K., {Russell}, H.~R., {et~al.} 2020, \aap, 633,
  A42

\bibitem[{{Sanders} {et~al.}(2016{\natexlab{a}}){Sanders}, {Fabian}, {Russell},
  {Walker}, \& {Blundell}}]{san16a}
{Sanders}, J.~S., {Fabian}, A.~C., {Russell}, H.~R., {Walker}, S.~A., \&
  {Blundell}, K.~M. 2016{\natexlab{a}}, \mnras, 460, 1898

\bibitem[{{Sanders} {et~al.}(2016{\natexlab{b}}){Sanders}, {Fabian}, {Taylor},
  {Russell}, {Blundell}, {Canning}, {Hlavacek-Larrondo}, {Walker}, \&
  {Grimes}}]{san16b}
{Sanders}, J.~S., {Fabian}, A.~C., {Taylor}, G.~B., {et~al.}
  2016{\natexlab{b}}, \mnras, 457, 82

\bibitem[{{Sauvageot} {et~al.}(2005){Sauvageot}, {Belsole}, \& {Pratt}}]{sau05}
{Sauvageot}, J.~L., {Belsole}, E., \& {Pratt}, G.~W. 2005, \aap, 444, 673

\bibitem[{{Seta} {et~al.}(2023){Seta}, {Federrath}, {Livingston}, \&
  {McClure-Griffiths}}]{set23}
{Seta}, A., {Federrath}, C., {Livingston}, J.~D., \& {McClure-Griffiths}, N.~M.
  2023, \mnras, 518, 919

\bibitem[{Shapiro \& Wilk(1965)}]{sha65}
Shapiro, S.~S. \& Wilk, M.~B. 1965, Biometrika, 52, 591

\bibitem[{{Simionescu} {et~al.}(2008){Simionescu}, {Werner}, {Finoguenov},
  {B{\"o}hringer}, \& {Br{\"u}ggen}}]{sim08}
{Simionescu}, A., {Werner}, N., {Finoguenov}, A., {B{\"o}hringer}, H., \&
  {Br{\"u}ggen}, M. 2008, \aap, 482, 97

\bibitem[{{Springel} {et~al.}(2018){Springel}, {Pakmor}, {Pillepich},
  {Weinberger}, {Nelson}, {Hernquist}, {Vogelsberger}, {Genel}, {Torrey},
  {Marinacci}, \& {Naiman}}]{spr18}
{Springel}, V., {Pakmor}, R., {Pillepich}, A., {et~al.} 2018, \mnras, 475, 676

\bibitem[{{Struble} \& {Rood}(1999)}]{str99}
{Struble}, M.~F. \& {Rood}, H.~J. 1999, \apjs, 125, 35

\bibitem[{{Str{\"u}der} {et~al.}(2001){Str{\"u}der}, {Briel}, {Dennerl},
  {Hartmann}, {Kendziorra}, {Meidinger}, {Pfeffermann}, {Reppin}, {Aschenbach},
  {Bornemann}, {Br{\"a}uninger}, {Burkert}, {Elender}, {Freyberg}, {Haberl},
  {Hartner}, {Heuschmann}, {Hippmann}, {Kastelic}, {Kemmer}, {Kettenring},
  {Kink}, {Krause}, {M{\"u}ller}, {Oppitz}, {Pietsch}, {Popp}, {Predehl},
  {Read}, {Stephan}, {St{\"o}tter}, {Tr{\"u}mper}, {Holl}, {Kemmer}, {Soltau},
  {St{\"o}tter}, {Weber}, {Weichert}, {von Zanthier}, {Carathanassis}, {Lutz},
  {Richter}, {Solc}, {B{\"o}ttcher}, {Kuster}, {Staubert}, {Abbey}, {Holland},
  {Turner}, {Balasini}, {Bignami}, {La Palombara}, {Villa}, {Buttler},
  {Gianini}, {Lain{\'e}}, {Lumb}, \& {Dhez}}]{str01}
{Str{\"u}der}, L., {Briel}, U., {Dennerl}, K., {et~al.} 2001, \aap, 365, L18

\bibitem[{Tamura {et~al.}(2014)Tamura, Yamasaki, Iizuka, Fukazawa, Hayashida,
  Ueda, Matsushita, Sato, Nakazawa, Ota, \& Takizawa}]{tam14}
Tamura, T., Yamasaki, N.~Y., Iizuka, R., {et~al.} 2014, 782, 38

\bibitem[{{Torrey} {et~al.}(2014){Torrey}, {Vogelsberger}, {Genel}, {Sijacki},
  {Springel}, \& {Hernquist}}]{tor14}
{Torrey}, P., {Vogelsberger}, M., {Genel}, S., {et~al.} 2014, \mnras, 438, 1985

\bibitem[{{Vogelsberger} {et~al.}(2013){Vogelsberger}, {Genel}, {Sijacki},
  {Torrey}, {Springel}, \& {Hernquist}}]{vog13}
{Vogelsberger}, M., {Genel}, S., {Sijacki}, D., {et~al.} 2013, \mnras, 436,
  3031

\bibitem[{{Vogelsberger} {et~al.}(2014{\natexlab{a}}){Vogelsberger}, {Genel},
  {Springel}, {Torrey}, {Sijacki}, {Xu}, {Snyder}, {Bird}, {Nelson}, \&
  {Hernquist}}]{vog14b}
{Vogelsberger}, M., {Genel}, S., {Springel}, V., {et~al.} 2014{\natexlab{a}},
  \nat, 509, 177

\bibitem[{{Vogelsberger} {et~al.}(2014{\natexlab{b}}){Vogelsberger}, {Genel},
  {Springel}, {Torrey}, {Sijacki}, {Xu}, {Snyder}, {Nelson}, \&
  {Hernquist}}]{vog14a}
{Vogelsberger}, M., {Genel}, S., {Springel}, V., {et~al.} 2014{\natexlab{b}},
  \mnras, 444, 1518

\bibitem[{{Weinberger} {et~al.}(2017){Weinberger}, {Springel}, {Hernquist},
  {Pillepich}, {Marinacci}, {Pakmor}, {Nelson}, {Genel}, {Vogelsberger},
  {Naiman}, \& {Torrey}}]{wei17}
{Weinberger}, R., {Springel}, V., {Hernquist}, L., {et~al.} 2017, \mnras, 465,
  3291

\bibitem[{{Wilms} {et~al.}(2000){Wilms}, {Allen}, \& {McCray}}]{wil00}
{Wilms}, J., {Allen}, A., \& {McCray}, R. 2000, \apj, 542, 914

\bibitem[{{Zhuravleva} {et~al.}(2014){Zhuravleva}, {Churazov}, {Schekochihin},
  {Allen}, {Ar{\'e}valo}, {Fabian}, {Forman}, {Sanders}, {Simionescu},
  {Sunyaev}, {Vikhlinin}, \& {Werner}}]{zhu14}
{Zhuravleva}, I., {Churazov}, E., {Schekochihin}, A.~A., {et~al.} 2014, \nat,
  515, 85

\bibitem[{{ZuHone} {et~al.}(2018){ZuHone}, {Miller}, {Bulbul}, \&
  {Zhuravleva}}]{zuh18}
{ZuHone}, J.~A., {Miller}, E.~D., {Bulbul}, E., \& {Zhuravleva}, I. 2018, \apj,
  853, 180

\bibitem[{{ZuHone} {et~al.}(2016){ZuHone}, {Miller}, {Simionescu}, \&
  {Bautz}}]{zuh16}
{ZuHone}, J.~A., {Miller}, E.~D., {Simionescu}, A., \& {Bautz}, M.~W. 2016,
  \apj, 821, 6

\end{thebibliography}

\end{document}